\documentclass[lettersize,journal]{IEEEtran}
\usepackage{amsmath,amsfonts}
\usepackage{algorithmic}
\usepackage{algorithm}
\usepackage{array}
\usepackage[caption=false,font=footnotesize,labelfont=rm,textfont=rm]{subfig}
\usepackage{textcomp}
\usepackage{stfloats}
\usepackage{graphicx}
\usepackage{epsfig,epsf,color,balance,cite}
\usepackage{epstopdf}
\usepackage{url}
\usepackage{verbatim}
\usepackage{graphicx}
\usepackage{cite}
\usepackage{amsmath}
\usepackage{amssymb}
\usepackage{bm}
\usepackage{color}
\allowdisplaybreaks[3]
\allowdisplaybreaks
\title{Rotatable Antenna Enabled Wireless Communication: Modeling and Optimization}

\author{Beixiong Zheng,~\IEEEmembership{Senior Member,~IEEE}, Qingjie Wu, Tiantian Ma, and Rui Zhang,~\IEEEmembership{Fellow,~IEEE}
\thanks{
The work of Beixiong Zheng is supported in part by the National Natural Science Foundation of China under Grant 62571193, the Guangdong program under Grant 2023QN10X446 and Grant 2023ZT10X148, and the GJYC program of Guangzhou under Grant 2024D01J0079 and Grant 2024D03J0006. This work of Rui Zhang is supported in part by National University of Singapore under Research Grants A-8003646-00-00 and A-8003676-00-00, and in part by the National Natural Science Foundation of China (No. 62331022), and the Guangdong Major Project of Basic and Applied Basic Research (No. 2023B0303000001).
\emph{(Corresponding author: Rui Zhang.)}
	
	B. Zheng, Q. Wu, and T. Ma are with the School of Microelectronics, South China University of Technology, Guangzhou 511442, China (e-mail: bxzheng@scut.edu.cn; miqjwu@mail.scut.edu.cn; mitiantianma@mail.scut.edu.cn).
	
	R. Zhang is with the Department of Electrical and Computer Engineering, National University of Singapore, Singapore 117583 (e-mail: elezhang@nus.edu.sg).}
}

\begin{document}
\markboth{IEEE Transactions on Communications, Vol. XX, No. XX, XXX 2026}{SKM: My IEEE article}
\maketitle

\begin{abstract}
%Fluid antenna system (FAS) and movable antenna (MA) have recently emerged as promising technologies to exploit new degrees of freedom (DoFs) in the spatial domain, which have attracted growing attention in wireless communication. In this paper, we propose a new rotatable antenna (RA) model to improve the performance of wireless communication systems.
In this paper, we propose a new rotatable antenna (RA) model to improve the performance of wireless communication systems. Different from conventional fixed antennas,
the proposed RA system can flexibly and independently alter the  boresight direction of each antenna via mechanical or electronic means to exploit new spatial degrees-of-freedom (DoFs).
Specifically, we investigate an RA-enabled uplink communication system, where the receive beamforming and the boresight directions of all RAs at the base station (BS) are jointly optimized to maximize the minimum signal-to-interference-plus-noise ratio (SINR) among all the users. In the special single-user and free-space propagation setup, the optimal boresight directions of RAs are derived in closed form with the maximum-ratio combining (MRC) beamformer applied at the BS.
%Moreover, we analyze the asymptotic performance with an infinite number of antennas based on this solution, which theoretically proves that the RA system can achieve a higher array gain than the fixed-antenna system.
In the general multi-user and multipath channel setup, we first propose an alternating optimization (AO) algorithm to alternately optimize the receive beamforming and the boresight directions of RAs in an iterative manner. Then, a two-stage algorithm that solves the formulated problem without the need for iteration is proposed to further reduce computational complexity.
Moreover, we extend the channel model to incorporate polarization effects and frequency-selective fading while catering to antenna boresight rotation.
Simulation results are provided to validate our analytical results and demonstrate that the proposed RA system can significantly improve the communication performance as compared to other benchmark schemes.
\end{abstract}

\vspace{-0.2cm}
\section{Introduction}
\IEEEPARstart{I}{n} the rapidly evolving landscape of global information and communications technology (ICT), the forthcoming sixth-generation (6G) wireless network is envisioned to support extremely high user/device density and a wider range of applications and services. These developments impose significantly higher performance requirements than those of previous generations~\cite{Wang2023On}.
Undoubtedly, multiple-input multiple-output (MIMO) is one of the most critical technologies to dramatically enhance the transmission rate and reliability of wireless networks.
%This is achieved through beamforming and multiplexing using multiple antennas at the transceivers~\cite{Bogale2016Massive}.}
However, the channel capacity and spectrum efficiency achieved by conventional MIMO are insufficient to meet the stringent requirements of 6G in its new applications. To further improve spatial resolution and degrees-of-freedom (DoFs), wireless networks tend to integrate drastically more antennas into arrays at the base stations (BSs), thereby evolving MIMO into massive MIMO, and ultimately into extremely large-scale MIMO~\cite{Lu2014An,Zheng2022A,Liu2023Near,Lu2024A,Bjornson2020Power}.

Although larger-scale MIMO can offer substantial array and spatial multiplexing gains, it comes at the expense of much higher hardware costs and power consumption. Furthermore, simply increasing the number of antennas cannot fully exploit the spatial DoFs, as traditional fixed antennas lack the flexibility to adjust their positions or orientations. Recently, fluid antenna system (FAS) and movable antenna (MA) have been proposed as promising technologies to overcome this limitation and have attracted growing attention in wireless communication~\cite{New2024A,Wong2021Fluid,Zhu2024Modeling}.
Compared to the fixed-antenna architecture, FAS/MA enables the local movement of antennas in a specified region through different antenna movement mechanisms. This flexibility enables proactive reshaping of the wireless channels into more favorable conditions, thereby achieving higher capacity without increasing the number of antennas.
Furthermore, leveraging the additional DoFs available at the physical layer, FAS/MA has been demonstrated to provide notable performance benefits in terms of interference mitigation, flexible beamforming, and multiplexing enhancement~\cite{Zhu2024Movable,Wong2022Fluid,New2024An}.
%By leveraging these capabilities of FAS/MA, substantial efforts have been devoted to integrating them with cutting-edge wireless technologies,

Building upon the aforementioned advantages, substantial efforts have been devoted to integrating FAS/MA with cutting-edge wireless technologies,
such as integrated sensing and communications (ISAC)~\cite{Qin2024Cramer}, intelligent reflecting surface (IRS)~\cite{Rostami2024On,Wei2025Movable,Zheng2025Intelligent}, and over-the-air computation~\cite{Zhang2024Fluid}. Nevertheless, although FAS/MA can bring numerous performance advantages, their practical implementation is highly constrained by the response~time and/or movement speed of the antennas. Additionally, existing works on FAS/MA still face limitations in terms of spatial flexibility and performance enhancement since only the positions of antennas are adjusted while their orientations are fixed.
To fully exploit all six-dimensional (6D) spatial DoFs, 6D movable antenna (6DMA) has been recently proposed to flexibly adjust both the three-dimensional (3D) position and 3D rotation of distributed antennas/arrays~\cite{Shao20256DTWC,Shao20256DJSAC,Shao2025Distributed,Shao2025Tutorial}. Based on the long-term/statistical user channel distribution, the 6DMA-equipped transceiver can allocate its antenna resources to improve the multi-user communication throughput.
%Additionally, 6DMA can achieve both array and geometric gains to enhance the performance of wireless sensing. 
%Although 6DMA provides a general model for position and rotation adjustable antennas, its implementation requires drastic changes of the current antenna architectures of existing BSs and thus may be practically cost-prohibitive. In addition, many new practical movement and rotational constraints should be considered in 6DMA systems~\cite{ref_6DMAmodel,ref_6DMAdiscrete,ref_6DMAMag}, making the joint design of positions and rotations of all 6DMA arrays highly challenging and sophisticated.

Motivated by the above, we propose in this paper a new antenna architecture, called rotatable antenna (RA), as a simplified realization of 6DMA to enhance the performance of wireless systems cost-effectively.
%In the RA system, the 3D orientation/boresight of each directional antenna can be independently adjusted, while its 3D position is kept constant to reduce the hardware cost and time/energy overhead associated with antenna position changes in 6DMA.
In the RA system, the 3D orientation/boresight of each directional antenna can be independently adjusted via mechanical or electronic means while its 3D position remains fixed.
%{\color{blue}As such, unlike the translation movement in MA and 6DMA, which requires additional space such as sliding tracks and movable area, RA only requires local rotational adjustment, which can be more readily achieved through compact mechanical or integrated electronic design~\cite{Zheng2025Rotatable, Xiong2025Intelligent}.
As shown in Table~\ref{tab:sum}, unlike MA and 6DMA that necessitate additional space for translational movement, RA only involves local rotational adjustment, which can be readily achieved through compact mechanical or integrated electronic design~\cite{Zheng2025Rotatable,zheng2026Tutorial,Xiong2025Intelligent}.
This thus offers RA great scalability and compatibility with existing wireless systems.
%This thus makes RA a streamlined and scalable solution that offers great compatibility with existing wireless systems.
%{\color{blue}Compared to the conventional fixed-antenna architecture, RA can enhance communication and sensing performance substantially by flexibly adjusting the antenna orientation/boresight.}
Compared to the fixed-antenna architecture, RA enhances communication and sensing performance by flexibly adjusting the antenna orientation/boresight.
%Furthermore, in a fixed-antenna-based MIMO system, the antenna array can only control the beam direction by imposing electrical phase differences among the antenna elements, and the array gain will be degraded when the beam direction deviates from the boresight direction of the antenna~\cite{ref_PhaseArray}. To overcome this drawback, RA is dedicated to exploiting the additional spatial DoFs in terms of the antenna orientation adjustment to actively align the main lobe direction of the antenna closer with the beam direction, while reconfiguring the directional gain pattern of the entire antenna array to optimize the allocation of radiation power.
%{\color{blue}In this way, RA provides a practical solution for enhancing array gains in desired directions to boost the transmit/receive signal power, while reducing the radiation power in undesired directions to avoid information leakage and interference.}
In this way, RA offers a practical and promising solution to enhance array gains towards desired directions while suppressing radiation towards undesired ones to reduce leakage and interference.
Therefore, by strategically designing the beamforming and antenna orientation/boresight, the RA system can be deployed to further improve the array/multiplexing gain and enhance the sensing resolution/range in various applications.
%, such as ISAC, massive machine-type communication (mMTC), simultaneous localization and mapping (SLAM), and others.

\begin{table*}[]
	\centering
		\caption{Comparison of Different Flexible Antenna Architectures~\cite{Zheng2025Rotatable}}
		\vspace{-0.3cm}
		\label{tab:sum}
		\renewcommand{\arraystretch}{1.5}
		\begin{tabular}{|c|c|c|c|c|}
			\hline
			\textbf{\begin{tabular}[c]{@{}c@{}}Antenna \\ Architecture\end{tabular}} &
			\textbf{Hardware  Implementation} &
			\textbf{\begin{tabular}[c]{@{}c@{}}Reconfigurable Parameter\end{tabular}} &
			\textbf{\begin{tabular}[c]{@{}c@{}}Deployment \\ Complexity\end{tabular}} &
			\textbf{\begin{tabular}[c]{@{}c@{}}System\\ Overhead\end{tabular}} \\ \hline\hline
			RA \cite{Zheng2025Rotatable} &
			\begin{tabular}[c]{@{}c@{}}Mechanical or electronic means\end{tabular} &
			\begin{tabular}[c]{@{}c@{}}Antenna 3D boresight direction\end{tabular} &
			Very low &
			Low \\ \hline
			FAS \cite{Wong2021Fluid} &
			Pixel antennas or liquid materials &
			\begin{tabular}[c]{@{}c@{}}Antenna shape and 3D position\end{tabular} &
			Moderate &
			Moderate \\ \hline
			MA \cite{Zhu2024Modeling} &
			Motors &
			Antenna 3D position &
			Moderate &
			Moderate \\ \hline
			6DMA \cite{Shao2025Tutorial} &
			Motors and flexible cables &
			\begin{tabular}[c]{@{}c@{}}Antenna 3D position and 3D rotation\end{tabular} &
			Moderate to high &
			Moderate to high \\ \hline
		\end{tabular}
	\vspace{-0.3cm}
\end{table*}

\begin{figure}[!t]  
\centering
	\includegraphics[width=3.0in]{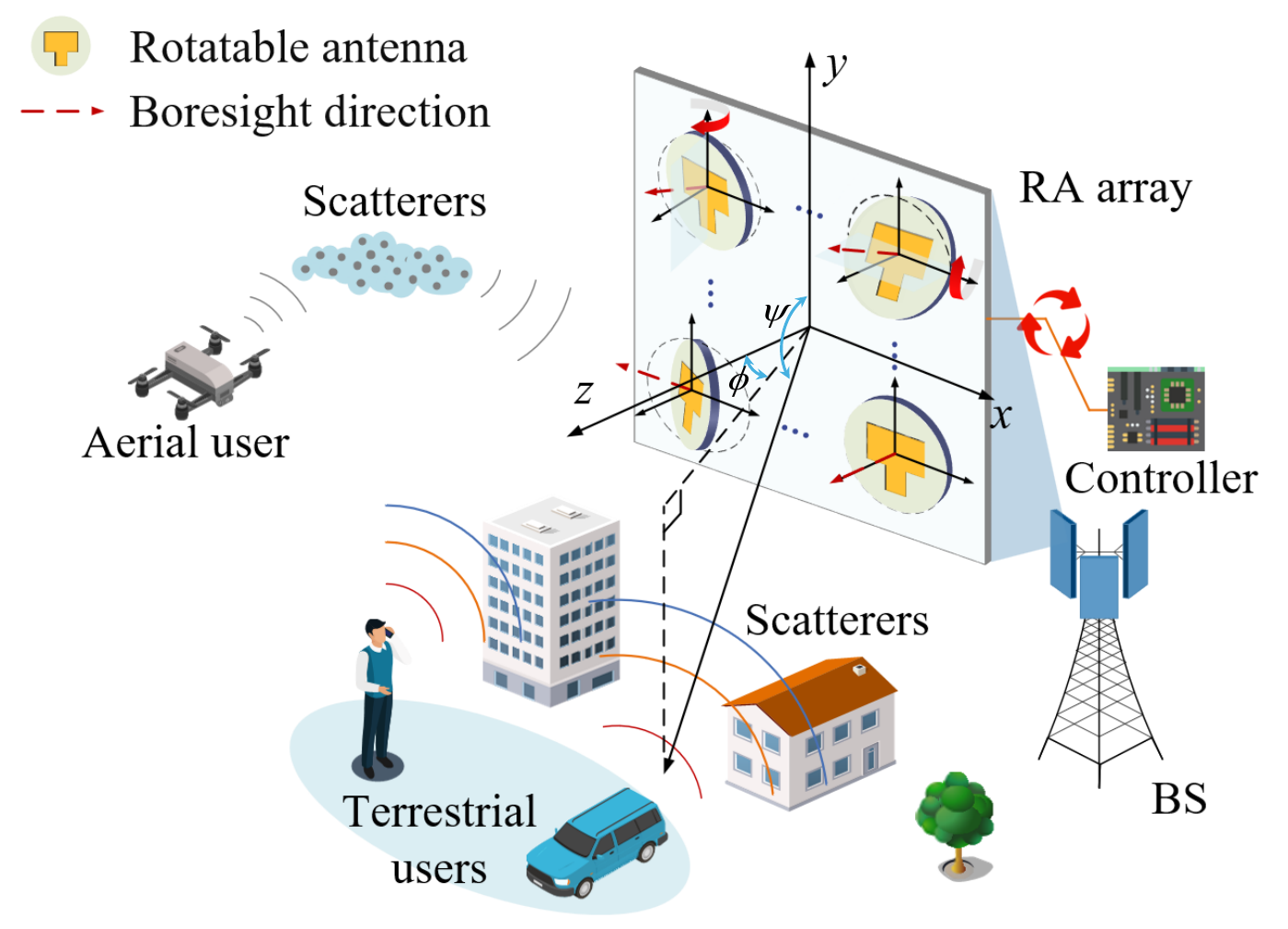} \vspace{-0.2cm}
	\caption{An RA-enabled uplink communication system.}
	\label{fig_system}\vspace{-0.5cm}
\end{figure}

Given the above technical advantages and potential applications of RA, we aim to investigate in this paper the  system modeling, performance analysis, and optimization algorithm design for an RA-enabled uplink communication system as shown in Fig.~\ref{fig_system}. The main contributions of this paper are summarized as follows:
\begin{itemize}
	\item{
%		Since the adjustment of antenna orientation/boresight should consider both the propagation environment and the antenna directional gain pattern, we first introduce a pointing vector to characterize the 3D boresight of each RA, and then present a new multipath geometric near-field channel model related to the pointing vectors of all RAs. 
%	{\color{blue}Adjusting the antenna orientation/boresight requires judicious consideration of both the propagation environment and the antenna's directional gain pattern. To facilitate this, we define a pointing vector to characterize the 3D orientation/boresight of each RA, and then construct a new multipath geometric near-field channel model based on the pointing vectors of all RAs.}
    We first introduce a pointing vector to characterize the 3D orientation/boresight direction of each RA, and incorporate the antenna boresight rotation into the multipath geometric-based channel model.
    The channel model is then further extended to account for polarization effects and frequency-selective fading while catering to antenna boresight rotation.}
%	Under this channel model, we further formulate a minimum signal-to-interference-plus-noise ratio (SINR) maximization problem to jointly optimize the receive beamforming and the pointing vectors of all RAs.}
	\item{For the special single-user and free-space propagation setup with the maximum-ratio combining (MRC) beamformer applied at the BS, we derive the optimal pointing vectors of RAs in closed form. Meanwhile, we also derive a closed-form expression and lower/upper bounds for the signal-to-noise ratio (SNR) under the uniform linear/planar array (ULA/UPA) setting, which show that the resultant SNR first increases linearly with the number of antennas and eventually converges to a certain limit.
	%{\color{blue}Additionally, we present an asymptotic analysis as the number of antennas goes to infinity, theoretically demonstrating that a wider rotational range for antenna boresight adjustment enables the proposed RA system to exploit more spatial DoFs, thereby achieving a higher array gain.}
	}
	\item{
%	For the general multi-user and multipath channel setup, to tackle the formulated min-SINR maximization problem for balancing the array directional gains among different users over their multipath channels, we first propose an alternating optimization (AO) algorithm that alternately optimizes the receive beamforming and the pointing vectors of RAs in an iterative manner until convergence is attained. 
	For the general multi-user and multipath channel setup, we formulate a minimum signal-to-interference-plus-noise ratio (SINR) maximization problem to balance the array directional gains among users. To address this problem, we 
	first propose an alternating optimization (AO) algorithm that alternately optimizes the receive beamforming and the pointing vectors of RAs in an iterative manner until convergence is achieved.}
	%In particular, with the optimal minimum mean-square error (MMSE) beamformer applied at the BS, the subproblem that optimizes the pointing vectors of RAs is solved by the successive convex approximation (SCA) technique. 
	To reduce computational complexity, we further propose a two-stage algorithm that solves a weighted channel power gain maximization problem based on the zero-forcing (ZF) beamformer without the need for iteration.
	\item{Simulation results validate our theoretical analysis and demonstrate that the proposed RA system can significantly improve communication performance over various benchmark schemes. It is shown that even with a small rotational range for antenna boresight adjustment, the RA system can achieve considerable performance gains over the fixed-antenna system. Furthermore, the performance advantages of RA in flexibly balancing the directional gain over multipath channels become more pronounced with stronger antenna directivity.
	}
\end{itemize}

The remainder of this paper is organized as follows. Section~\uppercase\expandafter{\romannumeral2} introduces the system model and problem formulation for designing the RA-enabled wireless communication system. In Section~\uppercase\expandafter{\romannumeral3}, we derive the optimal closed-form solution and analyze the asymptotic performance under the single-user setup. 
In Section~\uppercase\expandafter{\romannumeral4}, we propose the AO algorithm and the two-stage algorithm to solve the formulated problem under the multi-user setup.
Simulation results are presented in Section~\uppercase\expandafter{\romannumeral5}  to evaluate the performance of the proposed system and algorithms. Finally, we conclude the paper in Section \uppercase\expandafter{\romannumeral6}.

\textit{Notation:} Upper-case and lower-case boldface letters denote matrices and column vectors, respectively. Superscripts $(\cdot)^T$, $(\cdot)^{\ast}$, $(\cdot)^H$, and $(\cdot)^{-1}$ stand for the transpose, conjugate, Hermitian transpose, and matrix inversion operations, respectively. The sets of $a\times b$ dimensional complex and real matrices are denoted by $\mathbb{C}^{a\times b}$ and $\mathbb{R}^{a\times b}$, respectively. $\left \lfloor \cdot \right \rfloor$ is the floor function. $\mathcal{O}(\cdot)$ denotes the standard big-O notation. For a vector $\bm{x}$, $\|\bm{x}\|$ denotes its $\ell_2$-norm, $\mathrm{Re}\{\bm{x}\}$ denotes its real part, $\mathrm{diag}(\bm{x})$ returns a diagonal matrix with the elements in $\bm{x}$ on its main diagonal, and $[\bm{x}]_{a:b}$ denotes the subvector of $\bm{x}$ consisting of the elements from $a$ to $b$. For a matrix $\mathbf{X}$, $\mathrm{Tr}(\mathbf{X})$ and $\mathrm{rank}(\mathbf{X})$ denote its trace and rank, $[\mathbf{X}]_{a,b}$ denotes the $(a,b)$-th entry of matrix $\mathbf{X}$, $[\mathbf{X}]_{a:b,c:d}$ denotes the submatrix of $\mathbf{X}$ consisting of the elements located in rows $a$ to $b$ and columns $c$ to $d$, and $\mathbf{X}\succeq 0$ implies that $\mathbf{X}$ is positive semi-definite. $\mathbf{I}$ and $\mathbf{0}$ denote an identity matrix and an all-zero matrix, respectively, with appropriate dimensions. The distribution of a circularly symmetric complex Gaussian (CSCG) random vector with zero mean and covariance matrix $\bm{\Sigma}$ is denoted by $\mathcal{N}_c(\bm{0},\bm{\Sigma})$; and $\sim$ stands for ``distributed as''.
\vspace{-0.5cm}
\section{System Model and Problem Formulation}
As shown in Fig. \ref{fig_system}, we consider an RA-enabled uplink communication system, where $K$ users (each equipped with a single isotropic antenna) simultaneously transmit their signals in the same time-frequency resource block to a BS equipped with a UPA consisting of $N$ directional RAs. Without loss of generality, we assume that the UPA is placed on the $x$-$y$ plane of a 3D Cartesian coordinate system and centered at the origin with $N = N_x N_y$, where $N_x$ and $N_y$ denote the numbers of RAs along $x$- and $y$-axes, respectively. The separation between adjacent RAs is denoted by $\Delta$, and thus the entire UPA size can be expressed as $N_x\Delta \times N_y\Delta$. The physical size of each RA element is denoted by $\sqrt{A}\times \sqrt{A}$ with $\sqrt{A} \leq \Delta$, and we define $\xi \triangleq \frac{A}{\Delta^2} \leq 1$ as the array occupation ratio of the effective antenna aperture to the overall UPA region.

Assuming that both $N_x$ and $N_y$ are odd numbers for notational convenience, the reference position of RA~$n$, which is located at the $n_x$-th column and $n_y$-th row on the UPA, can be expressed as
\vspace{-0.2cm}\begin{align}
	\label{deqn_ex14a}
	\mathbf{w}_n \triangleq \mathbf{w}_{n_x,n_y} \triangleq \mathbf{w}_{(n_y-1)N_x+n_x} = [n_x\Delta,n_y\Delta,0]^T,
\end{align}
where $n_x = 0,\pm 1,\dots,\pm \frac{N_x-1}{2}$ and $n_y = 0,\pm 1,\dots,\pm \frac{N_y-1}{2}$. Let $r_k$ denote the distance between the center of the UPA and user~$k$ with $k = 1,2,\dots,K$. Accordingly, the position of user~$k$ is denoted by $\mathbf{u}_k = [r_k\Phi_k,r_k\Psi_k,r_k\Omega_k]^T$, with $\Phi_k \triangleq \operatorname{sin}\psi_k\operatorname{sin}\phi_k$, $\Psi_k \triangleq \operatorname{cos}\psi_k$, and $\Omega_k \triangleq \operatorname{sin}\psi_k\operatorname{cos}\phi_k$, where $\psi_k \in [0,\pi]$ and $\phi_k \in [-\frac{\pi}{2},\frac{\pi}{2}]$ denote the zenith and azimuth angles of user~$k$ with respect to the origin of the coordinate system, respectively. Accordingly, the distance between user~$k$ and RA~$n$ can be expressed as
\vspace{-0.2cm}\begin{align}
	\label{deqn_ex0a}
	r_{k,n}^{\mathrm{B-U}} & = \|\mathbf{u}_k - \mathbf{w}_n\| \nonumber\\
	& = r_k \sqrt{1 - 2n_x\delta_k\Phi_k - 2n_y\delta_k\Psi_k + (n_x^2 + n_y^2)\delta_k^2},
\end{align}
where $\delta_k \triangleq \frac{\Delta}{r_k}$ and $n = 1,2,\dots,N$. Note that $\delta_k \ll 1$ since the RA separation $\Delta$ is generally on the order of wavelength in practice.

\begin{figure}[!t]
	\centering
	\subfloat[Rotational angles]{
		\hspace{-0.4cm}\includegraphics[width=1.8in]{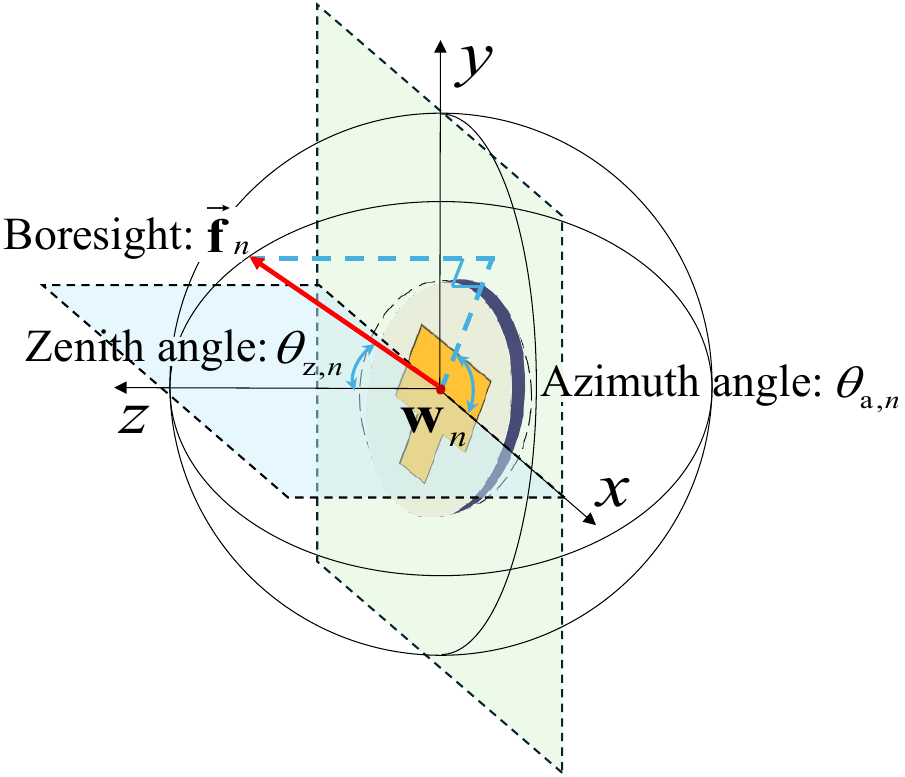}}\hspace{-0.25cm}
	\subfloat[Directional gain pattern]{
		\includegraphics[width=1.8in]{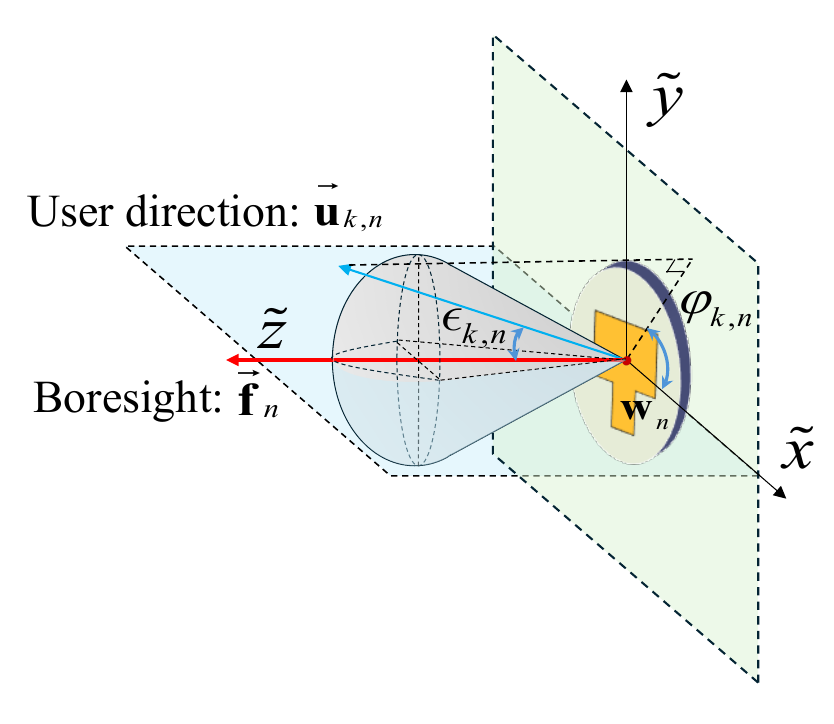}}\hspace{-0.4cm}
    \caption{Illustration of rotational angles and directional gain pattern of RA~$n$.}
	\label{fig_angle}\vspace{-0.4cm}
\end{figure}
\vspace{-0.4cm}
\subsection{Antenna Boresight Rotation}
The initial orientations/boresights of all RAs are assumed to be parallel to the $z$-axis, and each RA's boresight direction can be independently adjusted in 3D space by mechanical or electronic methods~\cite{Zheng2025Rotatable,zheng2026Tutorial,Xiong2025Intelligent}.
Specifically, mechanical control methods typically utilize a servo motor or micro-electromechanical system (MEMS) to physically steer the antenna boresight~\cite{Baek2003A}.
Note that MEMS-based actuators generally operate with milliwatt-level power consumption and millisecond-scale response times~\cite{Yang2025Low,Peng2020Monolithic}.
In contrast, electronic control methods maintain fixed antenna orientations while enabling the antenna boresight rotation through electronic techniques such as reconfigurable parasitic radiator element loading or PIN diode switching, achieving much faster response time at the microsecond scale~\cite{Zhang2022Highly,Towfiq2018A,Lotfi2017Printed,Shu2019An}.
%It is noted that mechanical control generally offers a wider range for boresight rotation, whereas electronic control provides much faster response and better compatibility with existing wireless systems.
%Notably, the power consumption and response speed of RAs depend on their specific implementation methods.
%In general, MEMS-based motors operate with power consumption in the milliwatt range and response times on the order of milliseconds~\cite{Yang2025Low,Peng2020Monolithic}.
%In comparison, electronically driven RAs offer much faster response times at microsecond scale~\cite{Zhang2022Highly,Zhang2024Dual}.}
%These characteristics highlight the practical feasibility of RA implementations and support the modeling assumptions adopted in this work.}

%As shown in Fig. \ref{fig_angle}(a), the 3D boresight adjustment of each RA can be described by a pair of \textit{deflection angles}: the zenith and azimuth angles with respect to the $z$-axis. Specifically, for RA~$n$, the zenith angle $\theta_{\mathrm{z},n}$ represents the angle between the boresight direction of RA~$n$ and the $z$-axis, while the azimuth angle $\theta_{\mathrm{a},n}$ is the angle between the projection of the boresight direction of RA~$n$ onto the $x$-$y$ plane and the $x$-axis. To characterize the 3D boresight direction of RA~$n$, we define a pointing vector determined by the zenith angle $\theta_{\mathrm{z},n}$ and the azimuth angle $\theta_{\mathrm{a},n}$ of RA~$n$ as
As shown in Fig.~\ref{fig_angle}(a), the 3D boresight direction of RA~$n$ can be characterized by a pointing vector, defined as
\vspace{-0.2cm}\begin{equation}
	\label{deqn_ex1a}
    \vec{\mathbf{f}}_n = \left[f_{x,n},f_{y,n},f_{z,n}\right]^T \in \mathbb{R}^{3\times 1},
\end{equation}
where $f_{x,n}$, $f_{y,n}$, and $f_{z,n}$ are the projections of RA~$n$'s pointing vector on the $x$-, $y$-, and $z$-axes, respectively. For the boresight rotation of RA~$n$, we let $\theta_{\mathrm{z},n}$ represent its zenith angle (i.e., the angle between the boresight direction of RA~$n$ and the $z$-axis) and $\theta_{\mathrm{a},n}$ represent its azimuth angle (i.e., the angle between the projection of the boresight direction of RA~$n$ onto the $x$-$y$ plane and the $x$-axis). Accordingly, we have $f_{x,n} = \operatorname{sin}\theta_{\mathrm{z},n}\operatorname{cos}\theta_{\mathrm{a},n}$, $f_{y,n} = \operatorname{sin}\theta_{\mathrm{z},n}\operatorname{sin}\theta_{\mathrm{a},n}$  and $f_{z,n} = \operatorname{cos}\theta_{\mathrm{z},n}$. Furthermore, we have $\|\vec{\mathbf{f}}_n\| = 1$ due to normalization.
To account for practical rotational constraint and mitigate antenna coupling between any two RAs, the zenith angle of each RA is confined to a specific range~\cite{Kumar2023Mutual}:
\vspace{-0.2cm}\begin{align}
	\label{deqn_ex2a}
	0 \leq \theta_{\mathrm{z},n} \leq \theta_{\mathrm{max}},\; \forall n,
\end{align}
where $\theta_{\mathrm{max}} \in [0,\frac{\pi}{2}]$ is the maximum zenith angle that each RA is allowed to adjust.%\footnote{\color{blue}To characterize the fundamental performance of the proposed RA system, we assume that the orientation/boresight of each RA can be continuously tuned in its rotational range, subject to the constraint in \eqref{deqn_ex2a}, while in practice it is usually chosen from a finite number of available directions for the ease of hardware implementation. The design of RA systems with discrete orientation/boresight rotation will be left for future work.}
\vspace{-0.4cm}
\subsection{Channel Model}
The effective antenna gain for each RA depends on both the signal arrival/departure angle and antenna directional gain pattern. In this paper, we consider the following widely used directional gain pattern for each RA (corresponding to the directional antenna with one narrow main lobe and negligible side lobes)~\cite{Balanis1996Antenna}
\vspace{-0.2cm}\begin{equation}
	\label{deqn_ex3a}
	{G_e(\epsilon,\varphi)} =
	\begin{cases}
		{G_0 \operatorname{cos}^{2p} (\epsilon),}&{\epsilon \in [0,\frac{\pi}{2}), \varphi \in [0,2\pi)} \\
		{0,}&{\text{otherwise},}
	\end{cases}
\end{equation}
where $(\epsilon,\varphi)$ is a pair of incident angles of the signal with respect to the RA's current boresight direction as shown in Fig. \ref{fig_angle}(b), $p\geq 0$ is the directivity factor that characterizes the beamwidth of the antenna's main lobe, and $G_0 = 2(2p + 1)$ is the maximum gain in the boresight direction (i.e., $\epsilon = 0$) that meets the law of power conservation. 

We consider the scattering environment with $Q$ distributed scatterer clusters, where the position of scatterer cluster~$q$ is represented by $\mathbf{c}_q \in \mathbb{R}^{3\times 1}$ with $q = 1,2,\dots,Q$. In the following, we consider the narrow-band frequency-flat channel model for ease of exposition. %{\footnote{\color{blue}The current narrow-band model can be extended to wideband scenarios \cite{Cui2024Near,Gao2021Wideband} by adopting orthogonal frequency division multiplexing (OFDM) technology, which divides the wideband channel into multiple orthogonal sub-bands, each undergoing frequency-flat fading. For this extension, the beamforming and orientations/boresights of RAs need to be jointly optimized across all sub-bands to maximize their achievable sum rate. {\color{blue} The corresponding simulation results for this wideband extension are presented in Section~\ref{sec:simulation}.}}}
The channel model will be extended to incorporate polarization effects and frequency-selective fading in Appendix E, with corresponding simulation results presented in Section~\ref{sec:simulation}-C.

Based on the \textit{Friis Transmission Equation} and the directional gain pattern adopted in \eqref{deqn_ex3a}, the channel power gain between user~$k$ and RA~$n$ can be modeled as~\cite{Friis1946A}
\vspace{-0.2cm}\begin{align}
	g_{\mathrm{U},k}(\vec{\mathbf{f}}_n) & \approx \int_{\mathbb{A}_n} \frac{1}{4\pi \|\mathbf{u}_k - \mathbf{a}\|^2} G_0 \left[\frac{\vec{\mathbf{f}}_n^T(\mathbf{u}_k - \mathbf{a})}{\|\mathbf{u}_k - \mathbf{a}\|}\right]_{+}^{2p} \mathrm{d}\mathbf{a} \nonumber\\
	& = \frac{A}{4\pi (r_{k,n}^{\mathrm{B-U}})^2} G_0 \left[\operatorname{cos}(\epsilon_{k,n})\right]_{+}^{2p}, \label{deqn_ex4a}
\end{align}
where $[x]_{+}\triangleq \operatorname{max}\{x,0\}$ denotes the positive operator, the integral space $\mathbb{A}_n = \left[n_x\Delta - \frac{\sqrt{A}}{2},n_x\Delta + \frac{\sqrt{A}}{2}\right] \times \left[n_y\Delta - \frac{\sqrt{A}}{2},n_y\Delta + \frac{\sqrt{A}}{2}\right]$ corresponds to the surface region of RA~$n$, $\mathbf{a}\in \mathbb{A}_n$ represents any point on plane $\mathbb{A}_n$, and $\operatorname{cos}(\epsilon_{k,n}) \triangleq \vec{\mathbf{f}}_n^T \vec{\mathbf{u}}_{k,n}$ is the projection between $\vec{\mathbf{f}}_n$ and $\vec{\mathbf{u}}_{k,n}$ with $\vec{\mathbf{u}}_{k,n} \triangleq \frac{\mathbf{u}_k - \mathbf{w}_n}{\|\mathbf{u}_k - \mathbf{w}_n\|}$ being the direction vector from RA~$n$ to user~$k$. Similarly, the channel power gain between scatterer cluster~$q$ and RA~$n$ is modeled as
\vspace{-0.2cm}\begin{align}
	g_{\mathrm{C},q}(\vec{\mathbf{f}}_n) \approx \frac{A}{4\pi (r_{q,n}^{\mathrm{B-C}})^2} G_0 \left[\operatorname{cos} (\tilde{\epsilon}_{q,n})\right]_{+}^{2p}, \label{deqn_ex5a}
\end{align}
where $r_{q,n}^{\mathrm{B-C}} = \|\mathbf{c}_q - \mathbf{w}_n\|$ is the distance between scatterer cluster~$q$ and RA~$n$, and $\operatorname{cos}(\tilde{\epsilon}_{q,n}) \triangleq \vec{\mathbf{f}}_n^T \vec{\mathbf{c}}_{q,n}$ is the projection between $\vec{\mathbf{f}}_n$ and $\vec{\mathbf{c}}_{q,n}$ with $\vec{\mathbf{c}}_{q,n} \triangleq \frac{\mathbf{c}_q - \mathbf{w}_n}{\|\mathbf{c}_q - \mathbf{w}_n\|}$ denoting the direction vector from RA~$n$ to scatterer cluster~$q$. Note that the channel power gains modeled in \eqref{deqn_ex4a} and \eqref{deqn_ex5a} account for both path gain and directional gain,
while the effect of polarization is incorporated in \eqref{deqn_ex3z} and \eqref{deqn_ex4z} of Appendix E.

For the multipath channel between RA~$n$ and user~$k$, by considering the geometric near-field propagation, the line-of-sight (LoS) channel component $h_{k}^{\mathrm{LoS}}(\vec{\mathbf{f}}_n)$ and the non-LoS (NLoS) channel component $h_{k}^{\mathrm{NLoS}}(\vec{\mathbf{f}}_n)$ can be separately modeled by~\cite{Lu2023Near,Dong2022Near}
\vspace{-0.2cm}\begin{align}
	h_{k}^{\mathrm{LoS}}(\vec{\mathbf{f}}_n) & = \sqrt{g_{\mathrm{U},k}(\vec{\mathbf{f}}_n)} e^{-j\frac{2\pi}{\lambda}r_{k,n}}, \label{deqn_ex8a}\\
	h_{k}^{\mathrm{NLoS}}(\vec{\mathbf{f}}_n) & = \sum_{q=1}^{Q}{\frac{\sqrt{\varsigma_q {g_{\mathrm{C},q}(\vec{\mathbf{f}}_n)}}}{r_{k,q}^{\mathrm{U-C}}}} e^{-j\frac{2\pi}{\lambda}(d_{q,n}+o_{k,q})+j\chi_q}, \label{deqn_ex9a}
\end{align}
where $\lambda$ is the signal wavelength, $\varsigma_q$ represents the radar cross section (RCS) of scatterer cluster~$q$, $\chi_q$ represents the phase shift introduced by scatterer cluster~$q$, and $r_{k,q}^{\mathrm{U-C}} = \|\mathbf{u}_k - \mathbf{c}_q\|$ denotes the distance between user~$k$ and scatterer cluster~$q$. Thus, by superimposing the LoS and NLoS channel components, the overall multipath channel between user~$k$ and the BS is given by\footnote{The current geometric channel model can be refined by incorporating the standardized path loss models defined in 3GPP TR 38.901~\cite{3gpp.38.901} and extended to account for frequency-selective fading, as shown in Appendix E.}
\vspace{-0.2cm}\begin{equation}
	\label{deqn_ex10a}
	\mathbf{h}_k(\mathbf{F}) = \mathbf{h}_k^{\mathrm{LoS}}(\mathbf{F}) + \mathbf{h}_k^{\mathrm{NLoS}}(\mathbf{F}),
\end{equation}
where $\mathbf{F} \triangleq [\vec{\mathbf{f}}_1,\vec{\mathbf{f}}_2,\dots,\vec{\mathbf{f}}_N] \in \mathbb{R}^{3\times N}$ is the pointing matrix of all RAs, $\mathbf{h}_k^{\mathrm{LoS}}(\mathbf{F}) \triangleq [h_{k}^{\mathrm{LoS}}(\vec{\mathbf{f}}_1),h_{k}^{\mathrm{LoS}}(\vec{\mathbf{f}}_2),\dots,h_{k}^{\mathrm{LoS}}(\vec{\mathbf{f}}_N)]^T$ and $\mathbf{h}_k^{\mathrm{NLoS}}(\mathbf{F}) \triangleq [h_{k}^{\mathrm{NLoS}}(\vec{\mathbf{f}}_1),h_{k}^{\mathrm{NLoS}}(\vec{\mathbf{f}}_2),\dots,h_{k}^{\mathrm{NLoS}}(\vec{\mathbf{f}}_N)]^T$ are the LoS and NLoS channel components from user~$k$ to the BS, respectively.
%It is observed that variations in $\mathbf{F}$ lead to the rotation of the directional gain pattern of each RA, thereby altering the effective multipath channel in~\eqref{deqn_ex10a}.
%% The orinigal sentence: The current geometric channel model can be further extended to incorporate more realistic propagation characteristics by adopting the standardized path loss models defined in 3GPP TR 38.901~\cite{3gpp.38.901} as well as extended to incorporate frequency-selective fading shown in Appendix E.

For the uplink communication, the received signal at the BS can be expressed as
\vspace{-0.2cm}\begin{equation}
	\label{deqn_ex11a}
	\mathbf{y} = \sum_{k=1}^{K}{\mathbf{h}_k(\mathbf{F}) \sqrt{P_k}s_k} + \mathbf{n},
\end{equation}
where $P_k$ and $s_k$ are the transmit power and information-bearing signal of user~$k$, respectively, and $\mathbf{n}$ is the additive white Gaussian noise (AWGN) vector, following the zero-mean CSCG distribution with variance $\sigma^2$, i.e., $\mathbf{n} \sim \mathcal{N}_c(\mathbf{0},\sigma^2\mathbf{I}_N)$. Upon receiving $\mathbf{y}$, the BS applies a linear receive beamforming vector $\mathbf{v}_k^H\in \mathbb{C}^{1\times N}$ with $\|\mathbf{v}_k\| = 1$ to extract the signal of user~$k$, i.e.,
\vspace{-0.2cm}\begin{equation}
	\label{deqn_ex12a}
	y_k = \mathbf{v}_k^{H} \mathbf{h}_k(\mathbf{F}) \sqrt{P_k}s_k + \sum_{j\ne k}{\mathbf{v}_k^{H} \mathbf{h}_j(\mathbf{F}) \sqrt{P_j}s_j}+ \mathbf{v}_k^{H}\mathbf{n}.
\end{equation}
Accordingly, the receive SINR at the BS for decoding the information from user~$k$ is given by
\vspace{-0.2cm}\begin{equation}
	\label{deqn_ex13a}
	\gamma_k = \frac{\bar{P}_k |\mathbf{v}_k^{H} \mathbf{h}_k(\mathbf{F})|^2}{\sum_{j\ne k}{\bar{P}_j|\mathbf{v}_k^{H} \mathbf{h}_j(\mathbf{F})|^2}+1},
\end{equation}
where $\bar{P}_k = \frac{P_k}{\sigma^2}$ denotes user~$k$'s equivalent transmit SNR.

%{\color{blue}\textit{Remark 1:} Although antenna orientation/boresight adjustments inevitably alter antenna polarization characteristics in practice, the proposed RA architecture introduces only minor variations in polarization characteristics due to its constrained rotational range at fixed antenna position and the absence of self-rotation.
%Moreover, the polarization ellipticity and handedness remain constant throughout the adjustment process, and thus efficient polarforming techniques~\cite{Zhou2024Polarforming,Shao2025Polarforming} can be applied to compensate the polarization induced channel variation such that the orientation/boresight adjustment by RA is still effective.
%Therefore, to focus on the effects of the directional gain pattern and explore the spatial DoFs offered by antenna orientation/boresight adjustments, polarization modeling is not considered in this paper for simplicity. Nevertheless, polarization modeling can be incorporated into future work on RA by integrating polarization matching efficiency into the channel power gain, following approaches similar to \cite{Zhou2024Polarforming} and \cite{Shao2025Polarforming}. Such extensions would require joint consideration of both the directional gain pattern and polarization matching during orientation/boresight optimization.}
\vspace{-0.30cm}
\subsection{Min-SINR Maximization Problem}
In this paper, we aim to maximize the minimum SINR among all the users by jointly optimizing the receive beamforming matrix $\mathbf{V}\triangleq [\mathbf{v}_1,\mathbf{v}_2,\dots,\mathbf{v}_K]$ and RA pointing matrix $\mathbf{F}$, subject to the rotational constraint given in \eqref{deqn_ex2a}. Thus, the min-SINR maximization problem is formulated as
\vspace{-0.25cm}\begin{subequations}\label{eq:1}
	\begin{alignat}{2}
		\text{(P1):} \quad \max_{\mathbf{V},\mathbf{F}} \quad & \min_{k} \gamma_k & \label{eq:1A}\\
		\mbox{s.t.} \quad 
		& 0\leq \operatorname{arccos}(\vec{\mathbf{f}}_n^T \mathbf{e}_3)\leq \theta_{\mathrm{max}},\; \forall n, \label{eq:1B}\\
		& \|\vec{\mathbf{f}}_n\| = 1,\; \forall n, \label{eq:1C}\\
		& \|\mathbf{v}_k\| = 1,\; \forall k, \label{eq:1D}
	\end{alignat}
\end{subequations}
where constraint \eqref{eq:1B} is equivalent to \eqref{deqn_ex2a} for ensuring that the rotation of boresight directions does not exceed the given range, and constraint \eqref{eq:1C} ensures that $\vec{\mathbf{f}}_n$ is a unit vector.

%In practice, channel estimation is necessary at the BS to obtain the required channel state information (CSI).
The antenna directional gain pattern in \eqref{deqn_ex3a} can be characterized and measured after antenna fabrication and thus treated as a given scaling factor over direction. As such, the BS only needs to acquire the key channel parameters such as path loss, angle of arrival/departure (AoA/AoD), and delay for optimizing RA boresight directions. Since adjusting the antenna orientation/boresight rather than its position does not alter the propagation geometry, existing geometry-based MIMO channel estimation methods for AoA/AoD and delay estimation, such as multiple signal classification (MUSIC) and compressed sensing (CS) approaches, remain applicable under antenna rotations. Then, least-squares (LS) or minimum mean-square error (MMSE)-based estimators can be applied to further estimate the path gain. By combining these estimated parameters with the pre-measured antenna directional gain pattern, the effective channel associated with each candidate boresight direction can be efficiently constructed, thereby enabling efficient RA boresight optimization without the need to estimate the channels with different boresight directions.
%{\color{blue} As demonstrated in \eqref{deqn_ex8a} and \eqref{deqn_ex9a}, rotating an antenna’s boresight or orientation does not change the propagation geometry, the antenna's radiation pattern, or fundamental channel parameters (e.g., path loss, angle of arrival/departure (AoA/AoD), and delay). Therefore, conventional MIMO channel estimation methods for fixed antennas remain applicable to the RA system~\cite{Zheng2025Rotatable}.}
In addition, an efficient channel estimation scheme tailored for RA systems was proposed in \cite{Xiong2025Efficient} to further improve estimation accuracy.
	
Denoting $\hat{\mathbf{h}}_k(\mathbf{F})$ as the estimated  channel state information (CSI) of the link from user~$k$ to the BS, the received signal is expressed as
\vspace{-0.25cm}\begin{equation}
	\label{deqn_ex14a}
	\mathbf{y} = \sum_{k=1}^{K}{\hat{\mathbf{h}}_k(\mathbf{F}) \sqrt{P_k}s_k} + \sum_{k=1}^{K}{\check{\mathbf{h}}_k\sqrt{P_k}s_k} + \mathbf{n},
\end{equation}where $\check{\mathbf{h}}_k \triangleq \mathbf{h}_k(\mathbf{F}) - \hat{\mathbf{h}}_k(\mathbf{F})$ stands for the CSI estimation error that is uncorrelated with the estimated channel $\hat{\mathbf{h}}_k(\mathbf{F})$.
In particular, when the number of scattering clusters is sufficiently large, both the channel and noise can be well approximated as jointly Gaussian processes.
If the CSI estimation error follows the zero-mean complex Gaussian distribution, i.e., $\sum_{k=1}^{K}{\check{\mathbf{h}}_k\sqrt{P_k}s_k}\sim \mathcal{N}_c\left(\mathbf{0},e^2\mathbf{I}_N\right)$ with $e$ accounting for the level of channel estimation error~\cite{Wang2007On}, the resultant error term can be incorporated into the AWGN term in \eqref{deqn_ex14a} for simplicity. 
Accordingly, the receive SINR for user~$k$ is given by
\vspace{-0.2cm}\begin{equation}
	\label{deqn_ex15a}
	\bar{\gamma}_k = \frac{\tilde{P}_k |\mathbf{v}_k^{H} \hat{\mathbf{h}}_k(\mathbf{F})|^2}{\sum_{j\ne k}{\tilde{P}_j|\mathbf{v}_k^{H} \hat{\mathbf{h}}_j(\mathbf{F})|^2}+1},
\end{equation}
where $\tilde{P}_k \triangleq \frac{P_k}{\sigma^2 + e^2}$ denotes user~$k$'s effective transmit SNR. It should be pointed out that $\bar{\gamma}_k$ in \eqref{deqn_ex15a} serves as a performance lower bound of the SINR modeled in \eqref{deqn_ex13a} due to the estimation error. Additionally, by carefully examining the SINR in \eqref{deqn_ex15a}, we find that it has a form similar to the SINR given in \eqref{deqn_ex13a}. % Therefore, the subsequent theoretical analysis and optimization algorithms developed based on problem (P1) can be readily applied to practical situations where the CSI estimation errors are taken into account.
Therefore, the subsequent theoretical analysis and optimization algorithms based on problem (P1) remain applicable in practical scenarios considering CSI estimation errors.

\vspace{-0.2cm}
\section{Single-User Case with Free-Space Propagation}
In this section, we consider the single-user and free-space propagation setup, i.e., $K = 1$ and $Q = 0$, to draw essential insights into (P1). Thus, the channel modeled in \eqref{deqn_ex10a} reduces to $\mathbf{h}_1(\mathbf{F}) = \mathbf{h}_1^{\mathrm{LoS}}(\mathbf{F})$. In this case, since no inter-user interference is present, problem (P1) is simplified to (by dropping the user index)
\vspace{-0.2cm}\begin{subequations}\label{eq:2}
	\begin{alignat}{2}
		\text{(P2):} \quad \max_{\mathbf{v},\mathbf{F}} \quad & \bar{P} |\mathbf{v}^{H} \mathbf{h}^{\mathrm{LoS}}(\mathbf{F})|^2 & \label{eq:2A}\\
		\mbox{s.t.} \quad 
		& \eqref{eq:1B}-\eqref{eq:1D}. \label{eq:2B}
	\end{alignat}
\end{subequations}
\vspace{-1.0cm}
\subsection{Optimal Closed-Form Solution}
For any given RA pointing matrix $\mathbf{F}$ in the single-user case, it is known that the MRC beamformer is the optimal receive beamforming solution to problem (P2)~\cite{Tse2005Fundamentals}, i.e., $\mathbf{v}_{\mathrm{MRC}} = \frac{\mathbf{h}^{\mathrm{LoS}}(\mathbf{F})}{\|\mathbf{h}^{\mathrm{LoS}}(\mathbf{F})\|}$. Thus, substituting $\mathbf{v}_{\mathrm{MRC}}$ into \eqref{eq:2A} yields the following SNR expression,
\vspace{-0.2cm}\begin{equation}
	\label{deqn_ex1b}
	\gamma = \bar{P} \|\mathbf{h}^{\mathrm{LoS}}(\mathbf{F})\|^2 = \frac{\bar{P}A}{4\pi}\sum_{n=1}^{N}{\frac{G_0\operatorname{cos}^{2p} (\epsilon_n)}{r_n^2}},
\end{equation}
where the superscript ``B-U'' of $r_{n}^{\mathrm{B-U}}$ is omitted in this section for simplicity. Exploiting the structure in \eqref{deqn_ex1b}, problem (P2) can be decomposed into $N$ subproblems, each of which independently optimizes the pointing vector of one RA. For RA~$n$, the corresponding subproblem is given by
\vspace{-0.2cm}\begin{subequations}\label{eq:3}
	\begin{alignat}{2}
		\text{(P3):} \quad \max_{\vec{\mathbf{f}}_n} \quad & \operatorname{cos}(\epsilon_n) = \vec{\mathbf{f}}_n^T \vec{\mathbf{u}}_n & \label{eq:3A}\\
		\mbox{s.t.} \quad 
		& 0\leq \operatorname{arccos}(\vec{\mathbf{f}}_n^T \mathbf{e}_3)\leq \theta_{\mathrm{max}}, \label{eq:3B}\\
		& \|\vec{\mathbf{f}}_n\| = 1, \label{eq:3C}
	\end{alignat}
\end{subequations}
where the constant term is omitted in \eqref{eq:3A}. By maximizing the projection between the unit vector $\vec{\mathbf{f}}_n$ and $\vec{\mathbf{u}}_n$, the optimal solution to problem (P3) is obtained as
\vspace{-0.2cm}\begin{equation}
	\label{deqn_ex2b}
	\vec{\mathbf{f}}_n^{\star} = \left[\operatorname{sin}\theta_{\mathrm{z},n}^{\star} \operatorname{cos}\theta_{\mathrm{a},n}^{\star}, \operatorname{sin}\theta_{\mathrm{z},n}^{\star} \operatorname{sin}\theta_{\mathrm{a},n}^{\star}, \operatorname{cos}\theta_{\mathrm{z},n}^{\star}\right]^T,
\end{equation}
where
\vspace{-0.2cm}\begin{subequations}\label{deqn_ex3b}
	\begin{align}
		\theta_{\mathrm{z},n}^{\star} & = \mathrm{min} 	\left \{\operatorname{arccos}\left(\vec{\mathbf{u}}_n^T \mathbf{e}_3\right),\theta_{\mathrm{max}} \right \}, \label{deqn_ex2b1}\\
		\theta_{\mathrm{a},n}^{\star} & = 	\operatorname{arctan2}\left(\vec{\mathbf{u}}_n^T \mathbf{e}_2, \vec{\mathbf{u}}_n^T \mathbf{e}_1 \right), \label{deqn_ex2b2}
	\end{align}
\end{subequations}
with $\mathbf{e}_1 \triangleq [1,0,0]^T$, $\mathbf{e}_2 \triangleq [0,1,0]^T$, and $\mathbf{e}_3 \triangleq [0,0,1]^T$.

According to the optimal pointing vector in \eqref{deqn_ex2b}, it can be inferred that each RA prefers to tune its antenna orientation/boresight towards the user. This is expected since the BS can achieve the maximum directional gain $NG_0$ when the boresight direction of each RA is aligned with the user direction, i.e., $\vec{\mathbf{f}}_n = \vec{\mathbf{u}}_n$. %To demonstrate the essential changes introduced by the proposed RA architecture, based on the directional gain pattern of each antenna element given in \eqref{deqn_ex3a}, we compare the array directional gain patterns of the RA and fixed-antenna arrays in Fig.~\ref{fig_single_arraygain}. For the RA array, the boresights of all RAs are adjusted to maximize the SNR in the user direction with $\psi = \frac{\pi}{2}$ and $\phi = \frac{\pi}{6}$ according to the optimal pointing vector in \eqref{deqn_ex2b}. It is observed that since RA array can focus its radiation power through aligning the main lobes of all antenna elements towards the desired direction, it significantly improves the array gain and narrows the beamwidth as compared to the fixed-antenna array. This result indicates that our proposed RA system has the ability to improve the communication performance by reconfiguring the array directional gain pattern according to different wireless environments and applications.
\vspace{-0.5cm}
\subsection{Asymptotic Performance Analysis}
\begin{figure}[!t] \centering
	\includegraphics[width=3in]{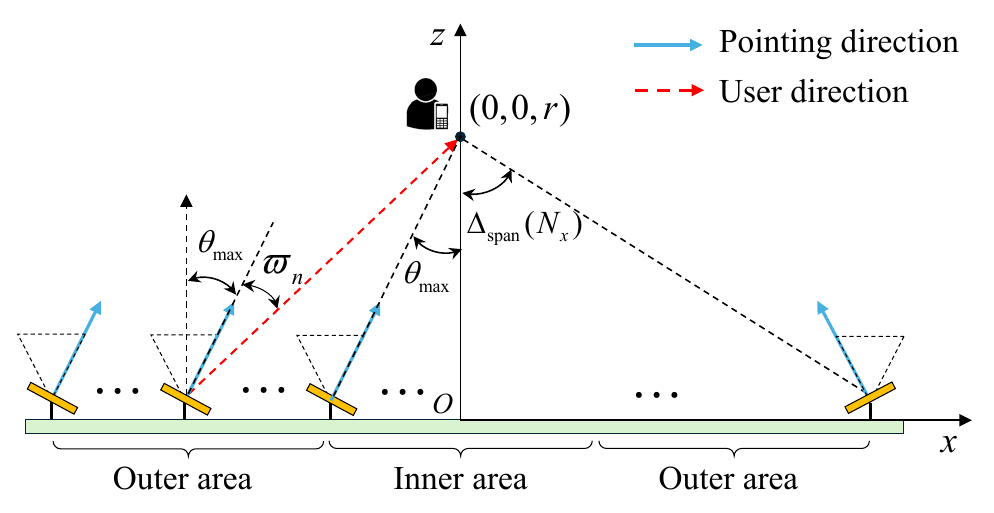}\vspace{-0.2cm}
	\caption{Illustration of the geometric relationship between the user and RAs.}
	\label{fig_asyangle}\vspace{-0.3cm}
\end{figure}

In this subsection, we focus on the performance analysis for the single-user system, and derive a closed-form expression for the SNR in the ULA case, lower/upper bounds for the SNR in the UPA case, and the asymptotic gains in both cases as the antenna number $N$ goes to infinity. For ease of exposition, we assume that the user is located along the $z$-axis (i.e., $\psi = \frac{\pi}{2}$ and $\phi = 0$) and its position is denoted by $\mathbf{u} = [0,0,r]^T$. In this case, based on the optimal pointing vector obtained in \eqref{deqn_ex2b}, the entire array region can be divided into inner and outer areas by comparing $\operatorname{arccos}(\vec{\mathbf{u}}_n^T \mathbf{e}_3)$ and $\theta_{\mathrm{max}}$, as shown in Fig.~\ref{fig_asyangle}. Specifically, if RA~$n$ is located in the inner area, we have $\operatorname{arccos}(\vec{\mathbf{u}}_n^T \mathbf{e}_3) \leq \theta_{\mathrm{max}}$ and $\varpi_n = 0$, i.e., RA~$n$ can adjust its boresight to perfectly align with the user direction to obtain the maximum antenna directional gain. Conversely, for RA~$n$ located in the outer area, we have $\operatorname{arccos}(\vec{\mathbf{u}}_n^T \mathbf{e}_3) > \theta_{\mathrm{max}}$ and $\varpi_n = \operatorname{arctan}\left(\sqrt{(n_x^2 + n_y^2)\delta^2}\right) - \theta_{\mathrm{max}}$, i.e., RA~$n$ can only serve the user with $\theta_{\mathrm{z},n} = \theta_{\mathrm{max}}$ and the RA boresight is offset from the user direction by an angle $\varpi_n > 0$ due to the zenith angle constraint in \eqref{deqn_ex2a}. Based on the above discussion, the projection between the user direction vector and the optimal pointing vector of RA~$n$ can be expressed as
\vspace{-0.2cm}\begin{align}
\hspace{-0.3cm}	\label{deqn_ex1c}
	\operatorname{cos}\varpi_n = \operatorname{cos}\left(\left[\operatorname{arctan}\left(\sqrt{(n_x^2 + n_y^2)\delta^2}\right) - \theta_{\mathrm{max}}\right]_{+}\right),
\end{align}
By substituting \eqref{deqn_ex1c} into \eqref{deqn_ex1b}, we can obtain the resultant SNR as \eqref{deqn_ex2c}, shown at the top of next page. The SNR in \eqref{deqn_ex2c} involves a double summation, which may make it difficult to gain useful insights. By approximating the double summation in \eqref{deqn_ex2c} as its corresponding double integral by leveraging $\delta \ll 1$ as in~\cite{Lu2022Communicating,Zheng2023Simultaneous,Feng2024Near}, the SNR can be rewritten in an integral form as \eqref{deqn_ex3c}, also shown at the top of next page.
\begin{figure*}[ht]
    \begin{align}
	    \label{deqn_ex2c}
	    \gamma = \frac{\bar{P}G_0 \xi \delta^2}{4\pi} \sum_{n_x=-\frac{N_x-1}{2}}^{\frac{N_x-1}{2}}{\sum_{n_y=-\frac{N_y-1}{2}}^{\frac{N_y-1}{2}}{\frac{\operatorname{cos}^{2p}\left(\left[\operatorname{arctan}\left(\sqrt{(n_x^2 + n_y^2)\delta^2}\right) - \theta_{\mathrm{max}}\right]_{+}\right)}{1 + (n_x^2 + n_y^2)\delta^2}}}.
    \end{align} \vspace{-0.6cm}
\end{figure*}

\begin{figure*}[ht]
	\vspace{-0.2cm} \begin{align}
		\label{deqn_ex3c}
		\gamma \simeq \frac{\bar{P}G_0 \xi \delta^2}{4\pi \Delta^2}\int_{-\frac{N_y\Delta}{2}}^{\frac{N_y\Delta}{2}}{\int_{-\frac{N_x\Delta}{2}}^{\frac{N_x\Delta}{2}}{\frac{\operatorname{cos}^{2p}\left(\left[\operatorname{arctan}\left(\sqrt{\frac{1}{r^2}(x^2 + y^2)}\right) - \theta_{\mathrm{max}}\right]_{+}\right)}{1 + \frac{1}{r^2}(x^2 + y^2)} \mathrm{d}x\mathrm{d}y }}.
	\end{align} \hrulefill \vspace{-0.4cm}
\end{figure*}

\textit{1) ULA-Based RA System:} To gain some insights, we first focus on the ULA setting. With $N_y = 1$ and $N_x = N$, the resultant SNR in \eqref{deqn_ex3c} reduces to
\vspace{-0.2cm}\begin{align}
	\label{deqn_ex4c}
	\hspace{-0.4cm}\gamma = \frac{\bar{P}G_0 \xi \delta^2}{4\pi \Delta}\int_{-\frac{N_x\Delta}{2}}^{\frac{N_x\Delta}{2}}{\frac{\operatorname{cos}^{2p}\left(\left[\operatorname{arctan}\left(|\frac{x}{r}|\right) - \theta_{\mathrm{max}}\right]_{+}\right)}{1 + \frac{x^2}{r^2}} \mathrm{d}x}.\hspace{-0.1cm}
\end{align}
It can be observed that the SNR given in \eqref{deqn_ex4c} is still very complicated for further analysis since the directivity factor $p$ exists as a power exponent of the cosine function. In the following, we discuss the typical case of $p = \frac{1}{2}$, i.e., the cosine pattern based on the projected aperture.

\textit{Theorem 1:} For the ULA-based RA system with cosine directional gain pattern (i.e., $p = \frac{1}{2}$) under the condition of $\delta \ll 1$, the maximum SNR achieved in the single-user setup can be expressed in closed-form as
\vspace{-0.2cm}\begin{equation}
	\label{deqn_ex5c}
	\hspace{-0.2cm}{\gamma} \hspace{-0.1cm}=\hspace{-0.1cm}
	\begin{cases}
		{\frac{2\bar{P}\xi\delta}{\pi} \triangle_{\mathrm{span}}(N_x),}&\hspace{-0.25cm}{N_x \leq \bar{N}_{x}} \\
		{\frac{2\bar{P}\xi\delta}{\pi}\left[\theta_{\mathrm{max}} + \operatorname{sin}\left(\triangle_{\mathrm{span}}(N_x) - \theta_{\mathrm{max}}\right) \right],}&\hspace{-0.25cm}{N_x > \bar{N}_{x},}
	\end{cases}\hspace{-0.2cm}
\end{equation}
where $\triangle_{\mathrm{span}}(N_x) \triangleq \operatorname{arctan}\left(\frac{N_x\delta}{2}\right)$ denotes the \textit{span angle} of the user, which is the angle formed by the two line segments connecting the user to the center and to one end of the RA array, as illustrated in Fig.~\ref{fig_asyangle}, and $\bar{N}_{x} \triangleq 2\left \lfloor  \frac{\operatorname{tan}\theta_{\mathrm{max}}}{\delta}\right \rfloor + 1$ is the maximum number of antennas in the inner area of the array.
\begin{IEEEproof}
	Please refer to Appendix A.
\end{IEEEproof}

Theorem~1 shows that with the applied MRC receive beamforming and the optimized RA pointing vectors, the resultant maximum SNR of the proposed ULA-based RA system scales with the antenna number $N_x$ according to the span angle $\triangle_{\mathrm{span}}(N_x)$. Furthermore, given the antenna size $A$, the antenna separation $\Delta$, and the propagation distance $r$, the maximum SNR of the ULA-based RA system depends on the allowable range for boresight rotation and the ULA size.

\textit{Remark 1:} By applying the linear approximation for the arctangent function, i.e., $\operatorname{arctan}(x) \approx \frac{\pi}{4}x,\;-1\leq x\leq 1$ \cite{Rajan2006Efficient}, the SNR in the first case of~\eqref{deqn_ex5c} can be approximated by $\gamma \approx \frac{1}{4}N_x\bar{P}\xi\delta^2$ since we have $0\leq \frac{N_x\delta}{2} \leq 1$ when $N_x \leq \bar{N}_{x}$. Thus, the resultant SNR increases linearly with the number of RAs when $N_x \leq \bar{N}_{x}$, i.e., $\triangle_{\mathrm{span}}(N_x) \leq \theta_{\mathrm{max}}$. 
It can be verified that $f(x) \triangleq \operatorname{sin}\left(\operatorname{arctan}\left(x\right) - \theta_{\mathrm{max}}\right)$ is a concave increasing function, and $\lim_{x\to \infty} f^{'}(x) = 0$.
This indicates that when $N_x > \bar{N}_{x}$, i.e., $\triangle_{\mathrm{span}}(N_x) > \theta_{\mathrm{max}}$, the growth rate of the maximum SNR gradually decreases as the number of RAs further increases, eventually approaching zero.

For the infinitely large-scale ULA such that $N_x \to \infty$, since $\operatorname{arctan}\left(\frac{N_x\delta}{2}\right) \to \frac{\pi}{2}$ as $\frac{N_x\delta}{2} \to \infty$, the SNR in \eqref{deqn_ex5c} reduces to
\vspace{-0.2cm}\begin{equation}
	\label{deqn_ex6c}
	\lim_{N_x \to \infty} \gamma = \frac{2\xi\delta}{\pi}\bar{P}\left(\theta_{\mathrm{max}} + \operatorname{cos}\theta_{\mathrm{max}}\right).
\end{equation}

It is observed that a higher asymptotic SNR can be achieved for a ULA-based RA system with a larger rotational range. By letting $\theta_{\mathrm{max}} = 0$ in \eqref{deqn_ex6c}, the asymptotic SNR for the conventional fixed-antenna system is given by
\vspace{-0.2cm}\begin{equation}
	\label{deqn_ex7c}
	\lim_{N_x \to \infty} \gamma_{\mathrm{FA}} = \frac{2\xi\delta}{\pi}\bar{P},
\end{equation}
where the boresight of each antenna is assumed to be parallel to the $z$-axis for ease of exposition.
Accordingly, the ratio of the asymptotic SNR of the RA system to that of the fixed-antenna system can be expressed as
\vspace{-0.2cm}\begin{equation}
	\label{deqn_ex13c}
	\frac{\lim_{N_x \to \infty} \gamma}{\lim_{N_x \to \infty} \gamma_{\mathrm{FA}}} = \theta_{\mathrm{max}} + \operatorname{cos}\theta_{\mathrm{max}} \geq 1,
\end{equation}
where the inequality holds since $f(x) = x + \operatorname{cos}x$ is a monotonically increasing function and $f(0) = 1$. By exploiting the additional spatial DoFs in terms of boresight rotation to improve the array gain, the proposed RA system with the optimal pointing vectors in \eqref{deqn_ex2b} will outperform the fixed-antenna system, and the performance gap increases with the allowable range of boresight rotation, as constrained by~\eqref{deqn_ex2a}.

\begin{figure}[!t]
	\centering
	\subfloat[Case 1]{
		\includegraphics[width=1.3in]{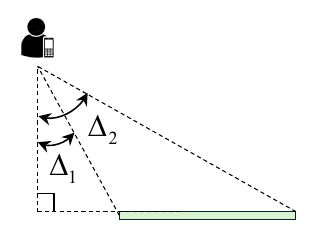}}
	\hspace{-0.3cm}\subfloat[Case 2]{
		\includegraphics[width=1in]{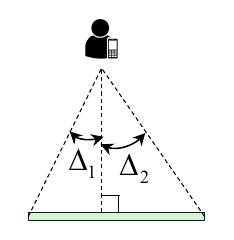}}
	\hspace{-0.25cm}\subfloat[Case 3]{
		\includegraphics[width=1.25in]{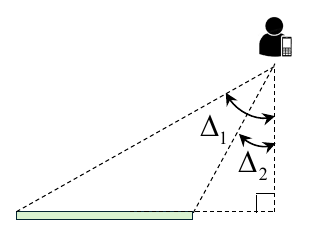}}
	\caption{Illustration of the geometric relationship between the user and ULA.}
	\label{fig_azimuth}\vspace{-0.3cm}
\end{figure}

\textit{Lemma 1:} If we define the closed-form SNR given in \eqref{deqn_ex5c} as a function with respect to the span angle of the user, i.e., $\gamma(\triangle_{\mathrm{span}}(N_x))$, Theorem~1 can be extended to the general case where the user is located around the ULA with $\psi = \frac{\pi}{2}$ and an arbitrary azimuth angle $\phi \in \left[-\frac{\pi}{2},\frac{\pi}{2}\right]$. Based on the three possible geometric relationships illustrated in Fig.~\ref{fig_azimuth}, the resultant SNR for a given azimuth angle $\phi$ is expressed as
\vspace{-0.2cm}\begin{equation}
	\label{deqn_ex14c}
	\hspace{-0.15cm}{\tilde{\gamma}}\hspace{-0.1cm} =\hspace{-0.1cm}
	\begin{cases}
		{\frac{1}{2\operatorname{cos}\phi}\left[\gamma(\triangle_2) - \gamma(\triangle_1)\right],}&\hspace{-0.25cm}{\text{Case 1:}\; \phi \in \hspace{-0.1cm}\left[-\frac{\pi}{2},-\frac{N_x\delta}{2}\right)} \\
		{\frac{1}{2\operatorname{cos}\phi}\left[\gamma(\triangle_1) + \gamma(\triangle_2)\right],}&\hspace{-0.25cm}{\text{Case 2:}\; \phi \in \hspace{-0.1cm}\left[-\frac{N_x\delta}{2},\frac{N_x\delta}{2}\right]} \\
		{\frac{1}{2\operatorname{cos}\phi}\left[\gamma(\triangle_1) - \gamma(\triangle_2)\right],}&\hspace{-0.25cm}{\text{Case 3:}\; \phi \in \hspace{-0.1cm}\left(\frac{N_x\delta}{2},\frac{\pi}{2}\right].}
	\end{cases}\hspace{-0.2cm}
\end{equation}
Since all geometric relationships shown in Fig.~\ref{fig_azimuth} will reduce to the symmetrical case in Fig.~\ref{fig_asyangle} when $N_x \to \infty$ regardless of $\phi$, the asymptotic SNR turns out to be the same as \eqref{deqn_ex6c}.

\textit{2) UPA-Based RA System:} Next, we consider the general UPA setting to analyze its asymptotic performance. For the UPA-based RA system with a moderate physical size, assuming that the entire UPA region is within the inner area, the following lemma yields an approximate SNR.

\textit{Lemma 2:} For a UPA-based RA system with $\sqrt{(N_x\Delta)^2 + (N_y\Delta)^2}\leq 2r\operatorname{tan}\theta_{\mathrm{max}}$ and $\theta_{\mathrm{max}} \leq \frac{\pi}{4}$, we have
\vspace{-0.2cm}\begin{align}
	\label{deqn_ex15c}
    \gamma \approx \frac{\bar{P}G_0\pi\xi\delta^2}{64}N_xN_y.
\end{align}

\begin{IEEEproof}
	Please refer to Appendix B.
\end{IEEEproof}

Lemma~2 shows that when the entire UPA region is located in the inner area, the optimal SNR in \eqref{deqn_ex15c} increases linearly with the antenna number $N = N_xN_y$. 

For a large RA array where the UPA region exceeds the inner area, it is challenging to obtain a closed-form expression for \eqref{deqn_ex3c} due to the double integral and the circular boundary between the inner and outer areas. Alternatively, we first derive its lower/upper-bounds for drawing useful insights as follows.

\textit{Theorem 2:} For the UPA-based RA system, defining $R_{\mathrm{lb}} = \frac{1}{2}\mathrm{min}\{N_x\Delta,N_y\Delta\}$ and $R_{\mathrm{ub}} = \frac{1}{2}\sqrt{(N_x\Delta)^2 + (N_y\Delta)^2}$ as the radii of the inscribed and circumscribed disks of the rectangular region $N_x\Delta \times N_y\Delta$ occupied by the UPA, the resultant SNR is lower/upper-bounded by
\vspace{-0.2cm}\begin{align}
	\label{deqn_ex8c}
	\mathcal{G}(R_{\mathrm{lb}},p,\theta_{\mathrm{max}}) \leq \gamma \leq \mathcal{G}(R_{\mathrm{ub}},p,\theta_{\mathrm{max}}),
\end{align}
where the function $\mathcal{G}(R,p,\theta_{\mathrm{max}})$ is defined as
\vspace{-0.2cm}\begin{align}
	\label{deqn_ex9c}
	 &\mathcal{G}(R,p,\theta_{\mathrm{max}}) = \frac{\bar{P} G_0 \xi}{2} \left[\frac{1}{2} \operatorname{ln}\left(1 + \left(\frac{D}{r}\right)^2 \right) + \right. \nonumber\\
	 &\left. \quad \quad \int_{\operatorname{arctan}\left(\frac{D}{r}\right) - \theta_{\mathrm{max}}}^{\operatorname{arctan}\left(\frac{R}{r}\right) - \theta_{\mathrm{max}}} \operatorname{cos}^{2p}\varpi\operatorname{tan}(\varpi + \theta_{\mathrm{max}}) \mathrm{d}\varpi\right],
\end{align}
with $D \triangleq \mathrm{min}\{R, r\operatorname{tan}\theta_{\mathrm{max}}\}$ being the radius of the inner area within which the RAs can adjust their boresights to precisely align with the user direction.

\begin{IEEEproof}
	Please refer to Appendix C.
\end{IEEEproof}

In \eqref{deqn_ex9c}, the first integral corresponds to the inner area with $\varpi_n = 0$, while the second integral corresponds to the outer area with $\varpi_n > 0$. The integral in \eqref{deqn_ex9c} is challenging to handle since the directivity factor $p$ exists as a power exponent of the cosine function. Similar to the ULA case, we focus on the cosine gain pattern (i.e., $p = \frac{1}{2}$) for convenience in the following discussion.

\textit{Lemma 3:} For $p = \frac{1}{2}$, the function $\mathcal{G}(R,p,\theta_{\mathrm{max}})$ can be expressed in closed form as \eqref{deqn_ex10c}, shown at the top of next page, where $\theta_R \triangleq \operatorname{arctan}\left(\frac{R}{r}\right)$.
\begin{figure*}[ht]
    \begin{align}
	\label{deqn_ex10c}
	\hspace{-0.1cm}{\mathcal{G}\left(R,\frac{1}{2},\theta_{\mathrm{max}}\right)} =
	\begin{cases}
		{\bar{P}\xi \operatorname{ln}\left(1 + \operatorname{tan}^2\theta_R \right),}&{R \leq r\operatorname{tan}\theta_{\mathrm{max}}} \\
		{2\bar{P}\xi \left[1 - \operatorname{ln}(\operatorname{cos}\theta_{\mathrm{max}}) - \operatorname{cos}(\theta_R - \theta_{\mathrm{max}}) +  \operatorname{sin}\theta_{\mathrm{max}} \operatorname{ln}\left(\frac{(1 + \operatorname{sin}\theta_R)(1 - \operatorname{sin}\theta_{\mathrm{max}})}{\operatorname{cos}\theta_R\operatorname{cos}\theta_{\mathrm{max}}}\right)\right],\hspace{-0.2cm}}&{R > r\operatorname{tan}\theta_{\mathrm{max}}}.
	\end{cases}
    \end{align} \hrulefill 
\end{figure*}

\begin{IEEEproof}
	Please refer to Appendix D.
\end{IEEEproof}

By combining Theorem~2 and Lemma~3, the lower/upper bounds for an RA system can be obtained according to the UPA size. To obtain the lower/upper bounds for the fixed-antenna system, we set $\theta_{\mathrm{max}} = 0$ in \eqref{deqn_ex9c} and calculate the integral in a manner similar to Appendix D, yielding
\vspace{-0.2cm}\begin{align}
	\label{deqn_ex11c}
	\mathcal{G}\left(R,\frac{1}{2},0\right) = 2\bar{P} \xi\left(1 - \operatorname{cos}\theta_R\right).
\end{align}
For the case of $\theta_{\mathrm{max}} > 0$, the inequality $\mathcal{G}(R,\frac{1}{2},\theta_{\mathrm{max}}) > 2\bar{P}\xi \left[1 - \operatorname{cos}\theta_R +  \operatorname{sin}\theta_{\mathrm{max}} \left(\operatorname{ln}\frac{1 + \operatorname{sin}\theta_R}{\operatorname{cos}\theta_R} + \operatorname{ln}\frac{1 - \operatorname{sin}\theta_{\mathrm{max}}}{1 - \operatorname{sin}^2\theta_{\mathrm{max}}}\right)\right] > \mathcal{G}(R,\frac{1}{2},0)$ always holds, indicating that the SNR's lower bound of the RA system is always higher than that of the fixed-antenna system when $R_{\mathrm{lb}} > r\operatorname{tan}\theta_{\mathrm{max}}$. As the UPA size $N_x\Delta,N_y\Delta \to \infty$, the radii of the inscribed and circumscribed disks of the UPA region $N_x\Delta \times N_y\Delta$ go to infinity, i.e., $R_{\mathrm{lb}},R_{\mathrm{ub}} \to \infty$. Therefore, the lower/upper bounds given by Theorem~2 approach the same limit due to the identical form of the function $\mathcal{G}(R,p,\theta_{\mathrm{max}})$ as $R \to \infty$. Based on the above, we conclude that for the UPA setting, the RA system also achieves a higher asymptotic SNR than the fixed-antenna system, as in the ULA case.
\vspace{-0.3cm}
\section{Multi-User Case under Multipath Channel}
In this section, we consider the general multi-user and multipath channel setup, i.e., $K > 1$ and $Q\geq 0$. Specifically, an AO algorithm and a two-stage algorithm are proposed to solve problem (P1) suboptimally, which offer different trade-offs between system performance and complexity.
\vspace{-0.4cm}
\subsection{AO Algorithm}
To overcome the challenges posed by the non-concavity~of the objective function in \eqref{eq:1A} and the intricate coupling between the receive beamforming vectors $\{\mathbf{v}_k\}$ and the RA pointing vectors $\{\vec{\mathbf{f}}_n\}$, an AO algorithm is proposed~to~alternately optimize the receive beamforming and RA pointing vectors in an iterative manner for the multi-user system.

\textit{1) Receive Beamforming Optimization:} For a given RA pointing matrix $\mathbf{F}$, the channel from user~$k$ to the BS modeled in \eqref{deqn_ex10a} becomes fixed. Accordingly, problem (P1) reduces to (by simplifying $\mathbf{h}_k(\mathbf{F})$ to $\mathbf{h}_k$)
\vspace{-0.2cm}\begin{subequations}\label{eq:4}
	\begin{alignat}{2}
		\text{(P4):} \quad \max_{\mathbf{V}} \quad & \min_{k} \quad \frac{\bar{P}_k |\mathbf{v}_k^{H} \mathbf{h}_k|^2}{\sum_{j\ne k}{\bar{P}_j|\mathbf{v}_k^{H} \mathbf{h}_j|^2}+1} & \label{eq:4A}\\
		\mbox{s.t.} \quad 
		& \eqref{eq:1C}. \label{eq:4B}
	\end{alignat}
\end{subequations}

The SINR in \eqref{eq:4A} is a generalized Rayleigh quotient with respect to $\mathbf{v}_k$, and thus the receive SINR for each user can be maximized by the minimum mean-square error (MMSE) beamforming \cite{Zheng2021Double}. Accordingly, the optimal solution to problem (P4) can be obtained as
\vspace{-0.2cm}\begin{equation}
	\label{deqn_ex1g}
	\mathbf{v}_k^{\mathrm{MMSE}} = \frac{\mathbf{C}_k^{-1} \mathbf{h}_k}{\|\mathbf{C}_k^{-1} \mathbf{h}_k\|},\; \forall k,
\end{equation}
%where $\mathbf{C}_k \triangleq \sum_{j\ne k}^{K}{\bar{P}_j \mathbf{h}_j \mathbf{h}_j^H} + \mathbf{I}_N$ is the interference-plus-noise covariance matrix.
where $\mathbf{C}_k \triangleq \sum_{j\ne k}^{K}{\bar{P}_j \mathbf{h}_j \mathbf{h}_j^H} + \mathbf{I}_N$ is the %{\color{blue}sampled interference-plus-noise covariance matrix estimated from samples under the quasi-stationary channel assumption.}
expected interference-plus-noise covariance matrix, which also corresponds to the sample covariance matrix over quasi-stationary channels.
To reduce the dimension of matrix inversion from $N\times N$ to $(K - 1)\times(K - 1)$, by applying the \textit{Woodbury matrix identity}, $\mathbf{C}_k^{-1}$ is equivalently expressed as
\vspace{-0.2cm}\begin{align}
	\label{deqn_ex2g}
	\mathbf{C}_k^{-1} = & (\mathbf{I}_N + \tilde{\mathbf{H}}_k \mathbf{P}_k \tilde{\mathbf{H}}_k^{H})^{-1} \nonumber\\
	= & \mathbf{I}_N - \tilde{\mathbf{H}}_k(\mathbf{P}_k^{-1} + \tilde{\mathbf{H}}_k^H\tilde{\mathbf{H}}_k)^{-1} \tilde{\mathbf{H}}_k^H,
\end{align}
where $\tilde{\mathbf{H}}_k \triangleq [\mathbf{h}_1,\dots,\mathbf{h}_{k-1},\mathbf{h}_{k+1},\dots,\mathbf{h}_K]$ and $\mathbf{P}_k \triangleq \operatorname{diag}\left(\bar{P}_1,\dots,\bar{P}_{k-1},\bar{P}_{k+1},\dots,\bar{P}_K\right)$.

\textit{2) RA Pointing Vector Optimization:} For a given receive beamforming $\mathbf{V}$, by introducing a slack variable $\eta$ to denote the minimum SINR, problem (P1) can be written as
\vspace{-0.2cm}\begin{subequations}\label{eq:5}
	\begin{alignat}{2}
		\text{(P5):} \quad \max_{\eta,\mathbf{F}} \quad & \eta & \label{eq:5A}\\
		\mbox{s.t.} \quad 
		& \gamma_k \geq \eta,\; \forall k, \label{eq:5B}\\
		& \eqref{eq:1B}, \eqref{eq:1C}. \label{eq:5C}
	\end{alignat}
\end{subequations}
Note that the above subproblem is still challenging to solve since constraints \eqref{eq:1C} and \eqref{eq:5B} are non-convex.

Based on \eqref{deqn_ex8a} and \eqref{deqn_ex9a}, the multipath channel between user~$k$ and RA~$n$ can be rewritten as
\vspace{-0.2cm}\begin{equation}
	\label{deqn_ex1d}
	h_{k}(\vec{\mathbf{f}}_n) = \alpha_{k,n} \left(\vec{\mathbf{f}}_n^T\vec{\mathbf{u}}_{k,n}\right)^p + \sum_{q=1}^{Q}{\beta_{k,n,q} \left(\vec{\mathbf{f}}_n^T\vec{\mathbf{c}}_{q,n}\right)^p},
\end{equation}
where
\vspace{-0.3cm}\begin{subequations}\label{deqn_ex10d}
	\begin{align}
		\hspace{-0.2cm}\alpha_{k,n} &\triangleq \frac{1}{r_{k,n}^{\mathrm{B-U}}} \sqrt{\frac{AG_0}{4\pi}} e^{-j\frac{2\pi}{\lambda}r_{k,n}^{\mathrm{B-U}}}, \label{deqn_ex10d1}\\
		\hspace{-0.2cm}\beta_{k,n,q} &\triangleq \frac{1}{r_{q,n}^{\mathrm{B-C}} r_{k,q}^{\mathrm{U-C}}} \frac{\sqrt{AG_0 \varsigma_q}}{4\pi} e^{-j\frac{2\pi}{\lambda}(r_{q,n}^{\mathrm{B-C}} + r_{k,q}^{\mathrm{U-C}})+j\chi_q}. \label{deqn_ex10d2}
	\end{align}
\end{subequations}
Then, we have $\mathbf{h}_k(\mathbf{F}) = [h_{k}(\vec{\mathbf{f}}_1),h_{k}(\vec{\mathbf{f}}_2),\dots,h_{k}(\vec{\mathbf{f}}_N)]^T$, and problem (P5) can be transformed into
\vspace{-0.2cm}\begin{subequations}\label{eq:6}
	\begin{alignat}{2}
		\text{(P6):} \quad \max_{\eta,\mathbf{F}} \quad & \eta & \label{eq:6A}\\
		\mbox{s.t.} \quad
		& \frac{\bar{P}_k |\mathbf{v}_k^{H} \mathbf{h}_k(\mathbf{F})|^2}{\sum_{j\ne k}{\bar{P}_j|\mathbf{v}_k^H \mathbf{h}_j(\mathbf{F})|^2} + 1} \geq \eta,\; \forall k, \label{eq:6B}\\
		& \operatorname{cos}(\theta_\mathrm{max}) \leq \vec{\mathbf{f}}_n^T \mathbf{e}_3 \leq 1,\; \forall n, \label{eq:6C}\\
		& \eqref{eq:1C}, \label{eq:6D}
	\end{alignat}
\end{subequations}
where constraint \eqref{eq:6C} is equivalent to \eqref{eq:1B}.

To deal with the fractional structure on the left-hand side of constraint \eqref{eq:6B}, we take the logarithmic operation on both sides of constraint \eqref{eq:6B}, resulting in an equivalent form for constraint \eqref{eq:6B}, i.e.,
\vspace{-0.3cm}\begin{align}
	\label{deqn_ex2d}
	 \hspace{-0.1cm}\operatorname{ln}\hspace{-0.1cm}\left(\bar{P}_k |\mathbf{v}_k^{H} \mathbf{h}_k(\mathbf{F})|^2\right) \hspace{-0.1cm}\geq\hspace{-0.1cm} \operatorname{ln}(\eta)
\hspace{-0.05cm}+\hspace{-0.05cm} \operatorname{ln}\hspace{-0.1cm}\left(\sum_{j\ne k}\hspace{-0.1cm}{\bar{P}_j|\mathbf{v}_k^H \mathbf{h}_j(\mathbf{F})|^2} \hspace{-0.1cm}+\hspace{-0.1cm} 1\right),\hspace{-0.1cm}
\end{align}
which is still difficult to handle since $h_{k}(\vec{\mathbf{f}}_n)$ in \eqref{deqn_ex1d} is neither convex nor concave due to the complex coefficients $\{\alpha_{k,n}\}$ and $\{\beta_{k,n,q}\}$. To tackle this challenge, we adopt the successive convex approximation (SCA) technique to approximate constraint \eqref{deqn_ex2d} as a convex constraint and obtain a local optimal solution to problem (P6) in an iterative manner. Without loss of generality, we present the procedure of the $(i+1)$-th iteration and denote the solutions of $\mathbf{F}$ and $\eta$ obtained in the $i$-th iteration by $\mathbf{F}^{(i)}$ and $\eta^{(i)}$, respectively. By using the first-order Taylor expansion at $\{\vec{\mathbf{f}}_n^{(i)}\}$, $|\mathbf{v}_k^{H} \mathbf{h}_k(\mathbf{F})|^2$ and $\operatorname{ln}\left(\sum_{j=1,j\ne k}^{K}{\bar{P}_j|\mathbf{v}_k^H \mathbf{h}_j(\mathbf{F})|^2} + 1\right)$ in \eqref{deqn_ex2d} can be respectively linearized as $\Lambda_k^{(i+1)}(\mathbf{F})$ and $\Gamma_k^{(i+1)}(\mathbf{F})$, shown at the top of the next page, where $\mathbf{h}_{k,n}^{'} \triangleq \frac{\partial h_{k}(\vec{\mathbf{f}}_n^{(i)})}{\partial \vec{\mathbf{f}}_n^{(i)}} = \tilde{\alpha}_{k,n} \vec{\mathbf{u}}_{k,n} + \sum_{q=1}^{Q}{\tilde{\beta}_{k,n,q} \vec{\mathbf{c}}_{q,n}}$ with $\tilde{\alpha}_{k,n} \triangleq \alpha_{k,n}p \left[(\vec{\mathbf{f}}_n^{(i)})^T\vec{\mathbf{u}}_{k,n}\right]_{+}^{p-1}$ and $\tilde{\beta}_{k,n,q} \triangleq \beta_{k,n,q}p \left[(\vec{\mathbf{f}}_n^{(i)})^T\vec{\mathbf{c}}_{q,n}\right]_{+}^{p-1}$. Similarly, an upper bound for $\operatorname{ln}(\eta)$ is obtained as $\Xi^{(i+1)}(\eta) \triangleq \operatorname{ln}(\eta^{(i)}) + \frac{\eta}{\eta^{(i)}} - 1$ by using its first-order Taylor expansions at $\eta^{(i)}$. In this way, constraint \eqref{deqn_ex2d} can be approximated by
\vspace{-0.2cm}\begin{figure*}[ht]
	\begin{align}
		\label{deqn_ex3d}
		\Lambda_k^{(i+1)}(\mathbf{F}) \triangleq |\mathbf{v}_k^{H} \mathbf{h}_k(\vec{\mathbf{f}}^{(i)})|^2 + \mathrm{Re}\left\{\left(\mathbf{v}_k^{H} \mathbf{h}_k(\vec{\mathbf{f}}^{(i)}) \right)^{\ast} \sum_{n=1}^{N}{ v_{k,n}^{\ast}\left(\mathbf{h}_{k,n}^{'}\right)^T(\vec{\mathbf{f}}_n - \vec{\mathbf{f}}_n^{(i)}) }\right\}.	
	\end{align}   \vspace{-0.4cm}
\end{figure*}
\begin{figure*}[ht]
	\vspace{-0.3cm} \begin{align}
		\label{deqn_ex4d}
		\hspace{-0.2cm}\Gamma_k^{(i+1)}(\mathbf{F}) \triangleq 
		\operatorname{ln}\left(\sum_{j=1,j\ne k}^{K}{\bar{P}_j|\mathbf{v}_k^H \mathbf{h}_j(\vec{\mathbf{f}}^{(i)})|^2} + 1 \right) + \frac{\sum_{j=1,j\ne k}^{K}{\bar{P}_j \mathrm{Re}\left\{\left(\mathbf{v}_k^{H} \mathbf{h}_j(\vec{\mathbf{f}}^{(i)}) \right)^{\ast} \sum_{n=1}^{N}{ v_{k,n}^{\ast}\left(\mathbf{h}_{j,n}^{'}\right)^T(\vec{\mathbf{f}}_n - \vec{\mathbf{f}}_n^{(i)})}\right\}}}{\sum_{j=1,j\ne k}^{K}{\bar{P}_j|\mathbf{v}_k^H \mathbf{h}_j(\vec{\mathbf{f}}^{(i)})|^2} + 1}.	
	\end{align} \hrulefill  \vspace{-0.4cm}
\end{figure*}
\begin{equation}
	\label{deqn_ex5d} \operatorname{ln}\left(\bar{P}_k\Lambda_k^{(i+1)}(\mathbf{F})\right) \geq \Gamma_k^{(i+1)}(\mathbf{F}) + \Xi^{(i+1)}(\eta),\; \forall k.
\end{equation}
Thus, problem (P6) can be approximated by the following problem in the $(i+1)$-th iteration.
\vspace{-0.2cm}\begin{subequations}\label{eq:7}
	\begin{alignat}{2}
		\text{(P7):} \quad \max_{\eta,\mathbf{F}} \quad & \eta & \label{eq:7A}\\
		\mbox{s.t.} \quad 
		& \eqref{eq:1C}, \eqref{eq:6C}, \eqref{deqn_ex5d}. \label{eq:7B}
	\end{alignat}
\end{subequations}
However, problem (P7) is still non-convex due to the unit-norm constraint for $\vec{\mathbf{f}}_n$ in \eqref{eq:1C}. For convenience, we first relax the equality constraint \eqref{eq:1C} as $\|\vec{\mathbf{f}}_n\| \leq 1$, yielding the following problem.
\vspace{-0.2cm}\begin{subequations}\label{eq:8}
	\begin{alignat}{2}
		\text{(P8):} \quad \max_{\eta,\mathbf{F}} \quad & \eta & \label{eq:8A}\\
		\mbox{s.t.} \quad 
		& \|\vec{\mathbf{f}}_n\| \leq 1,\; \forall n, \label{eq:8B}\\
		& \eqref{eq:6C}, \eqref{deqn_ex5d}. \label{eq:8C}
	\end{alignat}
\end{subequations}
It can be verified that problem (P8) is a convex optimization problem, which can be solved via the CVX solver~\cite{Boyd2004Convex}. Note that the optimal value obtained by problem (P8) serves as an upper bound for that of problem (P7) due to the relaxation of the equality constraint \eqref{eq:1C}.

\begin{algorithm}[!t]
	\caption{Proposed AO Algorithm for Solving (P1).}
	\begin{algorithmic}[1] \label{alg1}
		\STATE {Input:} Pointing vector $\mathbf{F}^{(0)}$, minimum receive SINR $\eta^{(0)}$, threshold $\varepsilon_2 > 0$, and maximum iteration number $I$.
		\STATE Initialization: $i \gets 0$.
		\REPEAT
		\STATE Given $\mathbf{F}^{(i)}$, calculate $\mathbf{V}^{(i+1)}$ according to \eqref{deqn_ex1g}.
		\STATE Given $\mathbf{V}^{(i+1)}$, $\mathbf{F}^{(i)}$, and $\eta^{(i)}$, obtain $\mathbf{F}^{(i+1)}$ and $\eta^{(i+1)}$ by solving problem (P8).
		\STATE Update $i=i+1$.
		\UNTIL $|\frac{\eta^{(i+1)} - \eta^{(i)}}{\eta^{(i)}}| \leq \varepsilon_2$ or $i > I$.
		\STATE {Output:} $\mathbf{V} = \mathbf{V}^{(i)}$ and $\mathbf{F} = \mathbf{F}^{(i)}$.
	\end{algorithmic} 
\end{algorithm} 

\textit{3) Overall Algorithm:}
Based on the results presented in the previous two subproblems, we propose the AO algorithm for problem (P1) by applying the block coordinate descent (BCD) method in Algorithm~\ref{alg1}. Specifically, all optimization variables in the original problem (P1) are partitioned into two blocks, i.e., $\{\mathbf{V},\mathbf{F}\}$. Then, the receive beamforming $\mathbf{V}$ and RA pointing matrix $\mathbf{F}$ are alternately optimized, by calculating \eqref{deqn_ex1g} and solving problem (P8), respectively, while keeping the other block of variables fixed. Furthermore, the obtained solution in each iteration is used as the input for the next iteration. 
The computational complexity of the proposed AO algorithm is mainly due to the matrix inversion for calculating the receive beamforming and the use of CVX solver to optimize the RA pointing matrix, which are performed in an iterative manner.
%The computational complexity of the proposed AO algorithm primarily depends on the iterative matrix inversion for receive beamforming and the use of a CVX solver to optimize the RA pointing matrix.
Therefore, the complexity order of Algorithm~\ref{alg1} is $\mathcal{O}\left(I\left(KN^3 + N^{3.5}\operatorname{ln}(1/\varepsilon_1)\right)\right)$, where $\varepsilon_1$ is the given solution accuracy and $I$ denotes the required iteration number for algorithm convergence.

Denote the objective value of problem (P1) based on a feasible solution $\{\mathbf{V},\mathbf{F}\}$ as $\eta(\mathbf{V},\mathbf{F})$. In step 4 of Algorithm~\ref{alg1}, for a given $\mathbf{F}^{(i)}$, the optimal solution of (P4) is obtained using the MMSE beamformer, which guarantees that $\eta(\mathbf{V}^{(i)},\mathbf{F}^{(i)}) \leq \eta(\mathbf{V}^{(i+1)},\mathbf{F}^{(i)})$. Subsequently, in step 5, given $\mathbf{V}^{(i+1)}$ and $\mathbf{F}^{(i)}$, problem (P8) is solved optimally to yield a high-quality solution for problem (P5), thereby ensuring that $\eta(\mathbf{V}^{(i+1)},\mathbf{F}^{(i)}) \leq \eta(\mathbf{V}^{(i+1)},\mathbf{F}^{(i+1)})$. As a result, we have $\eta(\mathbf{V}^{(i)},\mathbf{F}^{(i)}) \leq \eta(\mathbf{V}^{(i+1)},\mathbf{F}^{(i+1)})$, which indicates that the objective function is non-decreasing over iterations. Since the objective is upper-bounded by a finite value, the AO algorithm is guaranteed to converge to a stationary point. 

From a feasibility perspective, Algorithm~\ref{alg1} solves the relaxed problem where the equality constraint $\|\vec{\mathbf{f}}_n\| = 1$ in the original problem (P1) is relaxed to the inequality constraint $\|\vec{\mathbf{f}}_n\| \leq 1$. Thus, in the solution obtained by Algorithm~\ref{alg1}, if the pointing vector $\vec{\mathbf{f}}_n$ is unit-modulus, i.e., the equality in \eqref{eq:1C} holds, then the relaxation is tight and the obtained solution is feasible to problem (P1). Otherwise, the pointing vector needs to be reconstructed as a unit vector based on the solution obtained by Algorithm~\ref{alg1}, i.e., $\vec{\mathbf{f}}_n^{\star} = \frac{\vec{\mathbf{f}}_n}{\|\vec{\mathbf{f}}_n\|}$.
\vspace{-0.4cm}
\subsection{Two-Stage Algorithm}
In this subsection, we propose another low-complexity algorithm, namely the two-stage algorithm, to solve problem (P1) without the need for iteration. Specifically, the pointing vectors of all RAs are optimized using the semidefinite relaxation (SDR) technique in the first stage, and the corresponding beamforming vector is obtained by the ZF beamformer in the second stage.

As observed in Section \uppercase\expandafter{\romannumeral4}-A, the main difficulty in optimizing the pointing vectors lies in the intricate power function structure as shown in \eqref{deqn_ex1d} and the fractional structure of the SINR. According to the law of power conservation, as the directivity factor $p$ increases, the maximum antenna gain $G_0$ in the boresight direction becomes larger and the antenna main lobe becomes narrower. Nevertheless, the variation in parameter $p$ does not change the relative magnitude relationship of radiation power in different directions. As such, we consider a typical value of $p$ to eliminate the power function structure. Specifically, for the cosine-square gain pattern with $p = 1$, the multipath channel between user~$k$ and RA~$n$ modeled in \eqref{deqn_ex1d} is expressed as the following linear combination form,
\vspace{-0.2cm}\begin{equation}
	\label{deqn_ex4e}
	\bar{h}_{k}(\vec{\mathbf{f}}_n) = \vec{\mathbf{f}}_n^T\mathbf{m}_{k,n},
\end{equation}
where $\mathbf{m}_{k,n} \triangleq \alpha_{k,n}\vec{\mathbf{u}}_{k,n} + \sum_{q=1}^{Q}{\beta_{k,n,q}\vec{\mathbf{c}}_{q,n}}$. For computational simplicity, the positive operators in \eqref{deqn_ex4a} and \eqref{deqn_ex5a} are omitted in \eqref{deqn_ex4e} since all the users and scatterer clusters are confined to the front half-space of the BS.

The ZF receive beamforming is then adopted to completely remove the inter-user interference, which requires $N \geq K$. Therefore, by applying the ZF receive beamforming, the SINR reduces to an SNR without inter-user interference. For user~$k$, the ZF receive beamforming, denoted by $\mathbf{v}_k^{\mathrm{ZF}}$, should satisfy $(\mathbf{v}_k^{\mathrm{ZF}})^H \bar{\mathbf{H}}_k = \mathbf{0}_{1\times (K-1)}$, where $\bar{\mathbf{H}}_k \triangleq [\bar{\mathbf{h}}_1,\dots,\bar{\mathbf{h}}_{k-1},\bar{\mathbf{h}}_{k+1},\dots,\bar{\mathbf{h}}_K]$ with $\bar{\mathbf{h}}_k \triangleq [\bar{h}_{k}(\vec{\mathbf{f}}_1),\bar{h}_{k}(\vec{\mathbf{f}}_2),\dots,\bar{h}_{k}(\vec{\mathbf{f}}_N)]^T$. Therefore, the ZF receive beamforming for user~$k$ is expressed as
\vspace{-0.2cm}\begin{equation}
	\label{deqn_ex1e}
	\mathbf{v}_k^{\mathrm{ZF}} = \frac{(\mathbf{I}_N - \bar{\mathbf{H}}_k(\bar{\mathbf{H}}_k^H \bar{\mathbf{H}}_k)^{-1}\bar{\mathbf{H}}_k^H) \bar{\mathbf{h}}_k}{\|(\mathbf{I}_N - \bar{\mathbf{H}}_k(\bar{\mathbf{H}}_k^H \bar{\mathbf{H}}_k)^{-1}\bar{\mathbf{H}}_k^H) \bar{\mathbf{h}}_k\|},\; \forall k,
\end{equation}
where $\mathbf{I}_N - \bar{\mathbf{H}}_k(\bar{\mathbf{H}}_k^H \bar{\mathbf{H}}_k)^{-1}\bar{\mathbf{H}}_k^H$ is the projection matrix into the space orthogonal to the columns of $\bar{\mathbf{H}}_k$. By substituting \eqref{deqn_ex1e} into \eqref{deqn_ex13a}, the resultant SNR for user~$k$ with ZF beamforming is given by
\vspace{-0.2cm}\begin{align}
	\label{deqn_ex2e}
	\gamma_{\mathrm{ZF},k} = \bar{P}_k\|\bar{\mathbf{h}}_k\|^2 \left(1 - \rho_{\mathrm{ZF},k}\right),
\end{align}
where
\vspace{-0.2cm}\begin{align}
	\label{deqn_ex3e}
	\rho_{\mathrm{ZF},k} = \frac{\bar{\mathbf{h}}_k^H \bar{\mathbf{H}}_k(\bar{\mathbf{H}}_k^H \bar{\mathbf{H}}_k)^{-1}\bar{\mathbf{H}}_k^H \bar{\mathbf{h}}_k}{\|\bar{\mathbf{h}}_k\|^2}
\end{align}
with $0 \leq \rho_{\mathrm{ZF},k} \leq 1$ denoting the SNR loss caused by the cancellation of inter-user interference with ZF beamforming. According to \eqref{deqn_ex2e}, the resultant SNR for user~$k$ based on ZF beamforming mainly depends on channel power gain $\|\bar{\mathbf{h}}_k\|^2$ when the SNR loss $\rho_{\mathrm{ZF},k}$ is given by a reasonable value.

\textit{1) First stage:} Let $\mathbf{m}_k \triangleq [\mathbf{m}_{k,1}^T,\mathbf{m}_{k,2}^T,\dots,\mathbf{m}_{k,N}^T]^T \in \mathbb{C}^{3N\times 1}$ and $\mathbf{f}\triangleq [\vec{\mathbf{f}}_1^T,\vec{\mathbf{f}}_2^T,\dots,\vec{\mathbf{f}}_N^T]^T \in \mathbb{R}^{3N\times 1}$. We formulate the following problem to optimize the pointing vectors.
\vspace{-0.2cm}\begin{subequations}\label{eq:9}
	\begin{alignat}{2}
		\text{(P9):} \quad \max_{\omega,\mathbf{f}} \quad & \omega & \label{eq:9A}\\
		\mbox{s.t.} \quad 
		& \rho_k\bar{P}_k\|\mathbf{f}^T \mathbf{m}_k\|^2\geq \omega,\; \forall k, \label{eq:9B}\\
		& \mathbf{f}_{3(n-1)+1:3n}^T \mathbf{e}_3 \geq \operatorname{cos}\theta_{\mathrm{max}},\; \forall n, \label{eq:9C}\\
		& \|\mathbf{f}_{3(n-1)+1:3n}\|^2 = 1,\; \forall n, \label{eq:9D}
	\end{alignat}
\end{subequations}
where $\rho_k$ can be initially set as $\rho_k = 1 - \rho_{\mathrm{ZF},k}$ based on \eqref{deqn_ex3e} with $\vec{\mathbf{f}}_n = \mathbf{e}_3,\;\forall n$, serving as a weight factor for user~$k$'s channel power gain, and constraints \eqref{eq:9C} and \eqref{eq:9D} are equivalent to \eqref{eq:6C} and \eqref{eq:6D}, respectively. Note that $\|\mathbf{f}^T \mathbf{m}_k\|^2 = \mathbf{f}^T \mathbf{M}_k \mathbf{f} = \mathrm{Tr}(\mathbf{M}_k\mathbf{f}\mathbf{f}^T)$ with $\mathbf{M}_k = \mathbf{m}_k \mathbf{m}_k^H$. Define $\bar{\mathbf{F}} = \mathbf{f}\mathbf{f}^T$, which needs to satisfy $\bar{\mathbf{F}} \succeq \mathbf{0}$ and $\mathrm{rank(\bar{\mathbf{F}})} = 1$. Since the rank-one constraint is non-convex, we apply semidefinite relaxation (SDR) technique to relax this constraint. As a result, problem (P9) is transformed into
\vspace{-0.2cm}\begin{subequations}\label{eq:10}
	\begin{alignat}{2}
		\hspace{-0.15cm}\text{(P10):} \quad \max_{\omega,\bar{\mathbf{F}}} \quad & \omega & \label{eq:10A}\\
		\mbox{s.t.} \quad 
		& \rho_k \bar{P}_k \mathrm{Tr}(\mathbf{M}_k\bar{\mathbf{F}}) \geq \omega,\; \forall k, \label{eq:10B}\\
		& \bar{\mathbf{F}}_{3(n-1)+1,3(n-1)+1} \geq \operatorname{cos}^2\theta_{\mathrm{max}},\; \forall n, \label{eq:10C}\\
		& \mathrm{Tr}(\bar{\mathbf{F}}_{3(n-1)+1:3n,3(n-1)+1:3n}) = 1, \forall n, \label{eq:10D}\\
		& \bar{\mathbf{F}} \succeq \mathbf{0}. \label{eq:10E}
	\end{alignat}
\end{subequations}
As problem (P10) is a convex semidefinite program (SDP), it can be optimally solved by the CVX solver with a complexity order of $\mathcal{O}((3N)^{3.5})$~\cite{Luo2010Semidefinite}, and we represent the optimal solution of problem (P10) as $\left\{\omega^{\star},\bar{\mathbf{F}}^{\star}\right\}$.

Since problem (P10) may not lead to a rank-one solution of $\bar{\mathbf{F}}$, the optimal objective value of problem (P10) serves as an upper bound of problem (P9). Thus, a rank-one approximation on $\bar{\mathbf{F}}^{\star}$ should be executed as an additional step to construct a feasible solution to problem (P9). If $\bar{\mathbf{F}}^{\star}$ is rank-one, we have $\bar{\mathbf{F}}^{\star} = \mathbf{f}^{\star} \mathbf{f}^{\star T}$, and $\mathbf{f}^{\star}$ will be a feasible and optimal solution to problem (P9). On the other hand, if the rank of $\bar{\mathbf{F}}^{\star}$ is larger than one, we define $\bar{\mathbf{f}} = \sqrt{\lambda_{\mathrm{max}}}\bm{\upsilon}_{\mathrm{max}}$, with $\lambda_{\mathrm{max}}$ and $\bm{\upsilon}_{\mathrm{max}}$ denoting the maximum eigenvalue and its corresponding eigenvector obtained through eigenvalue decomposition of $\bar{\mathbf{F}}^{\star}$, respectively. This serves as our candidate solution to problem (P9) since the best rank-one approximation to $\bar{\mathbf{F}}^{\star}$ is given by $\mathbf{f}^{\star} = \lambda_{\mathrm{max}}\bm{\upsilon}_{\mathrm{max}}\bm{\upsilon}_{\mathrm{max}}^T$ \cite{Luo2010Semidefinite}.

\textit{2) Second Stage:} After obtaining the stacked pointing vector $\mathbf{f}^{\star}$ in the first stage and letting $\vec{\mathbf{f}}_n = \mathbf{f}_{3(n-1)+1:3n}^{\star},\;\forall n$, we reconstruct the channels according to \eqref{deqn_ex8a}--\eqref{deqn_ex10a}. Then, the corresponding ZF beamforming can be calculated by \eqref{deqn_ex1e}.

The computational complexity of the proposed two-stage algorithm is mainly due to the SDP and singular value decomposition in the first stage, and the calculation of the ZF beamforming in the second stage. Thus, the complexity order of the two-stage algorithm is given by $\mathcal{O}\left(KN^3 + (3N)^3 + (3N)^{3.5}\right)$. Since the two-stage algorithm only needs to solve problem (P10) and calculate the beamforming in \eqref{deqn_ex1e} for one time, it requires lower computational complexity than the AO algorithm.
%, especially when the number of antennas is large or the convergence accuracy of the AO algorithm is strict.

\textit{Remark 2:} Although the two-stage algorithm may experience some performance loss by setting $p = 1$, it has lower complexity without the need for iteration. In contrast, the AO algorithm, while having higher computational complexity, is applicable to any value of $p$. It is worth noting that both the AO and two-stage algorithms can also be applied to the single-user setup with arbitrary $Q$ scatterer clusters by replacing the MMSE/ZF receive beamforming of \eqref{deqn_ex1g}/\eqref{deqn_ex1e}  with MRC beamforming. Nevertheless, for the single-user setup with $Q = 0$, since the AO and two-stage algorithms can only achieve suboptimal solutions and result in higher computational complexity, they are much less efficient than the optimal closed-form solution derived in \eqref{deqn_ex2b} of Section \uppercase\expandafter{\romannumeral3}-A. On the other hand, although the original problem can be transformed into a pointing vector optimization problem by substituting the MMSE receive beamforming of \eqref{deqn_ex1g} into \eqref{deqn_ex13a}, it is found that the objective function becomes even more complicated due to the presence of matrix inversion, which is difficult to handle and thus not considered in this paper.
\vspace{-0.3cm}
\section{Simulation Results}
\label{sec:simulation}
In this section, we present simulation results to evaluate the performance of our proposed RA-enabled communication system as well as the optimization algorithms for the joint design of receive beamforming and RA pointing vectors. In the following simulations, similar to \cite{Shao20256DTWC}, we assume the system operates at 2.4 GHz with a wavelength of $\lambda =$ 0.125 meter (m), the receiver noise power is set to $\sigma^2 = -$80 dBm, the antenna separation is $\Delta = \frac{\lambda}{2}$, and the size of each antenna is $A = \frac{\lambda^2}{4\pi}$. Unless otherwise stated, a square UPA-based system with $N_x = N_y =$ 4 is considered, the transmit power of all users is set to the same value, i.e., $P_k = P =$ 10 dBm$,\; \forall k$, and the maximum zenith angle allowed for RA boresight rotation is set as $\theta_{\mathrm{max}} = \frac{\pi}{6}$.
\vspace{-0.3cm}
\subsection{Single-User System}
First, we consider a single-user system under free-space propagation, where the distance between the center of the array and the user is set to $r =$ 15 m. In this subsection, we consider the cosine gain pattern (i.e., $p = \frac{1}{2}$) to validate the performance analysis in Section \uppercase\expandafter{\romannumeral3}-B. The received signal power $P_R$ at the BS, which is proportional to the resultant SNR due to the relationship $P_R = \sigma^2 \gamma$, is considered as the performance metric.

\begin{figure}[!t] \centering
	\includegraphics[width=2.67in]{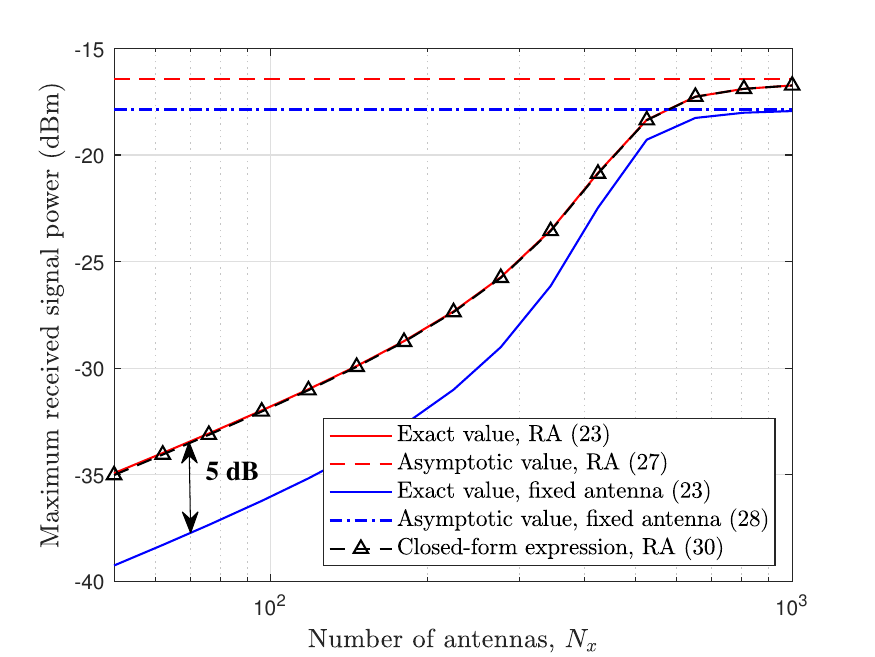}\vspace{-0.2cm}
	\caption{Maximum received signal power versus the number of antennas $N_x$ for the ULA system with $\phi = \frac{5\pi}{12}$ and $\psi = \frac{\pi}{2}$.}
	\label{fig_single_ULA}\vspace{-0.3cm}
\end{figure}

%To compare the directional gains of RA and fixed-antenna systems, 
Fig.~\ref{fig_single_ULA} plots the maximum received signal power versus the number of antennas $N_x$ for a ULA system. The asymptotic values given in \eqref{deqn_ex6c} and \eqref{deqn_ex7c} are also shown in the figure. It is first observed that the closed-form SNR derived in \eqref{deqn_ex14c} matches perfectly with the exact value calculated by \eqref{deqn_ex2c}, which validates the correctness of Theorem~1 and Lemma~1. Additionally, for a small to moderate number of antennas, the received signal powers of both the RA and fixed-antenna systems increase linearly with $N_x$, which is in accordance with Remark~1. However, as $N_x$ further increases, it is observed that the received signal powers of both RA and fixed-antenna systems eventually approach their asymptotic values. Meanwhile, the RA system reaches its asymptotic limit later and achieves up to 1.43 dB gain over the fixed-antenna system, which corroborates the accuracy of analytical result in \eqref{deqn_ex13c} as $10\mathrm{log}_{10}\frac{\lim_{N_x \to \infty} \tilde{\gamma}}{\lim_{N_x \to \infty} \tilde{\gamma}_{\mathrm{fixed}}} = 10\mathrm{log}_{10}\left(\frac{\pi}{6} + \operatorname{cos}\frac{\pi}{6}\right) \approx$~1.43~dB. Furthermore, when $N_x \leq 100$, the RA system achieves up to 5 dB gain over the conventional fixed-antenna system.
%This indicates that the proposed RA architecture can achieve a high array gain due to the additional spatial DoFs induced by the antenna boresight rotation.

\begin{figure}[!t] \centering
	\includegraphics[width=2.67in]{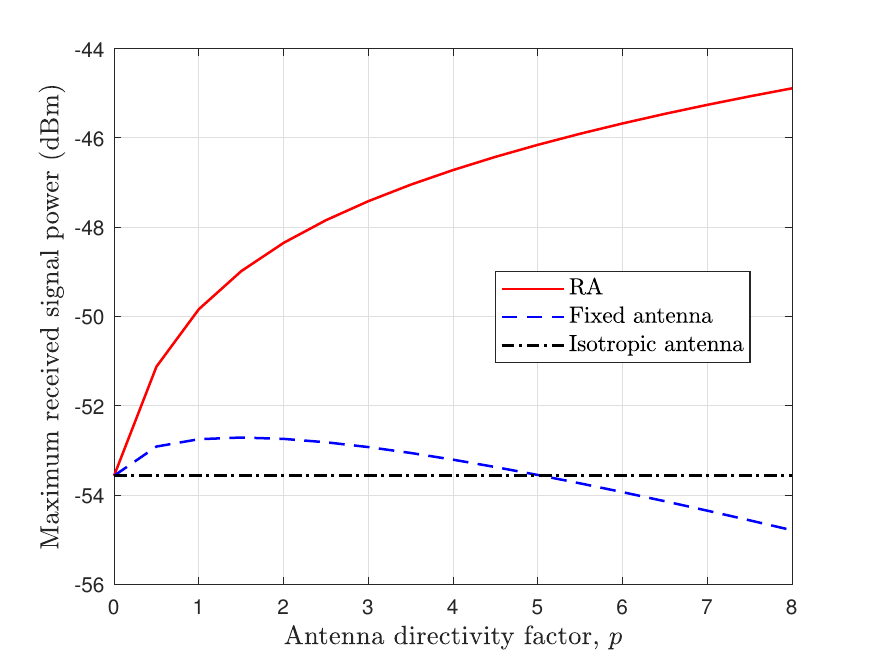}\vspace{-0.2cm}
	\caption{Maximum received signal power versus the antenna directivity factor~$p$ for the UPA system with $N_x = N_y = 4$ and $\psi = \frac{\pi}{2}$.}
	\label{fig_single_UPA}\vspace{-0.3cm}
\end{figure}

In Fig.~\ref{fig_single_UPA}, we plot the received signal power versus the antenna directivity factor $p$ for the UPA system. It is observed that the received signal power of the RA system increases as $p$ increases. This is expected since a higher directivity (i.e., larger $p$) results in greater radiation power in the boresight direction, leading to a higher array gain focused on the user direction. Conversely, the received signal power of the fixed-antenna system first increases and then decreases with $p$. This is expected since with a relatively low antenna directivity (e.g., $p \leq 4$), the main lobe is sufficiently broad to encompass the user direction, allowing the fixed antenna to offer enhanced directional gain compared to the isotropic antenna. However, as $p$ continues to increase, the main lobe becomes narrower, eventually excluding the user direction from its effective coverage. %Consequently, the directional gain in the user direction decreases with the further increase of $p$. 
Moreover, it is observed that the received signal power of the RA system is consistently higher than that of the fixed-antenna system regardless of $p$. This highlights the performance advantage of the RA system, which benefits from the radiation energy focusing capability.

\begin{figure}[!t] \centering
	\includegraphics[width=2.67in]{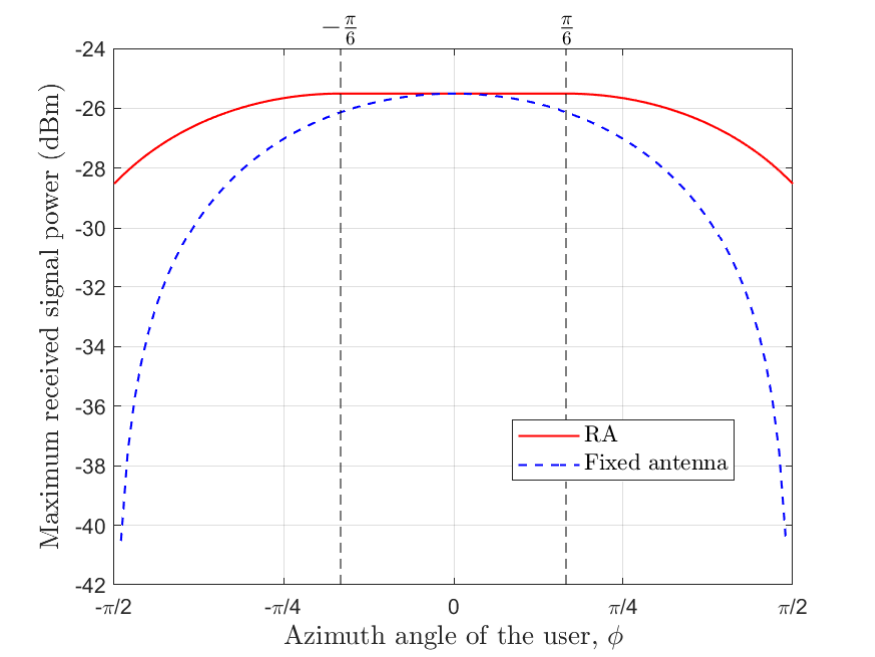}\vspace{-0.2cm}
	\caption{Maximum received signal power versus the azimuth angle of the user~$\phi$ for the UPA system with $N_x = N_y = 4$ and $\psi = \frac{\pi}{2}$.}
	\label{fig_single_angle}\vspace{-0.3cm}
\end{figure}

Fig. \ref{fig_single_angle} shows the received signal power versus the azimuth angle of the user $\phi$ for the UPA system, where the zenith angle of the user is given by $\psi = \frac{\pi}{2}$. As $\phi$ increases from 0 to $\frac{\pi}{2}$ or decreases from 0 to $-\frac{\pi}{2}$, the received signal power of the fixed-antenna system drastically decreases. This is due to the fact that the array directional gain pattern of the fixed-antenna system is fixed and the radiation power only focuses on the region directly in front of the array. In contrast, the RA array enables the BS to maintain more stable and uniform received signal power over the entire angular region.
%, particularly when $\phi$ lies within the interval $[-\frac{\pi}{6},\frac{\pi}{6}]$, where each RA can adjust its boresight within the allowable rotational range to precisely align with the user direction. 
%In addition, the RA system achieves much higher received signal power than the fixed-antenna system even when the user direction deviates significantly from the array's main direction, i.e., $\phi \to -\frac{\pi}{2}$ or $\frac{\pi}{2}$. 
These advantages stem from the RA array's ability to flexibly reconfigure its directional gain pattern to enhance the directional gain in the user direction.
%The above results indicate that RA array has the potential to uniformly enhance communication coverage performance in its front half-space.
\vspace{-0.3cm}
\subsection{Multi-User Narrow-Band System}\label{Multi-User Narrow-Band}
Next, we consider a multi-user system under the multipath channel model with $K = 4$ users and $Q = 8$ scatterer clusters. Specifically, four users are uniformly distributed in four distinct directions in front of the BS. The distance of each user to the BS is independently and randomly selected from a uniform distribution within $[30,50]$~m. Around these users, eight scatterer clusters are randomly distributed. In the following, we present simulation results by averaging over 500 independent channel realizations. Meanwhile, we aim to maximize the rate achievable by all $K$ users as the performance metric:
\vspace{-0.2cm}\begin{align}
	C = \min_{k}\; \mathrm{log}_2\left(1 + \gamma_k\right) = \mathrm{log}_2\left(1 + \min_{k}\;\gamma_k\right).
\end{align}
%where the weighting coefficient $\varrho$ with $0 \leq \varrho < 1$ is introduced to account for the system overhead.
%{\color{blue} where the overhead ratio $\varrho$ with $0 \leq \varrho < 1$ is introduced to account for the system overhead arising from the antenna rotation latency, CSI acquisition, and power consumption.
%In our simulations, the overhead ratio for the RA and fixed-antenna systems are set as $\varrho_{\mathrm{RA}} = 0.1$ and $\varrho_{\mathrm{FA}} = 0.05$, respectively.}}

\begin{figure}[!t] \centering
	\includegraphics[width=2.67in]{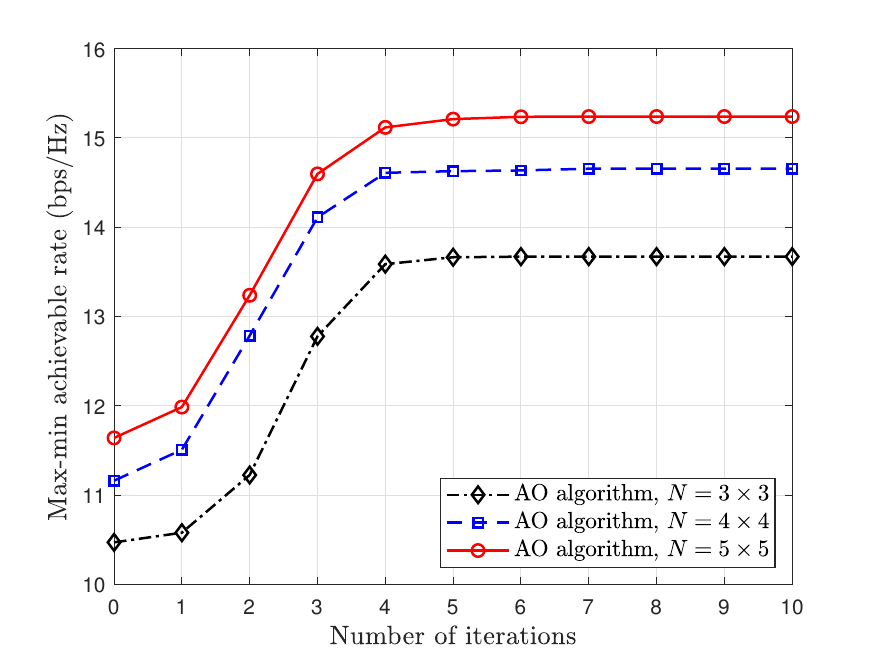}\vspace{-0.2cm}
	\caption{Convergence behavior of the proposed AO algorithm.}
	\label{fig_convergence}\vspace{-0.3cm}
\end{figure}

In Fig.~\ref{fig_convergence}, we show the convergence behavior of the proposed AO algorithm (i.e., Algorithm 1) with different numbers of RAs. It is observed that the max-min achievable rate increases over iterations and converges within six iterations. This indicates that the proposed AO algorithm converges quickly and is effective.
To further validate the performance advantages of our proposed RA system, we consider the following three benchmark schemes for comparison:
\begin{itemize}
	\item{\bf{Random orientation design}}: In this scheme, the orientation of each RA is randomly generated within the rotational ranges given by \eqref{deqn_ex2a}, and the MMSE receive beamforming is applied at the BS.
%	\item{\bf{Array-wise orientation adjustment}:} In this scheme, we adjust the orientation of the entire antenna array instead of that of each antenna element. {\color{blue}By applying the MMSE receive beamforming, the optimal orientation of the antenna array is obtained by using exhaustive search.}
	\item{\bf{Fixed orientation design}:} In this scheme, the orientations of all RAs are fixed at their reference orientations, i.e., $\vec{\mathbf{f}}_n = \mathbf{e}_3,\; \forall n$, and the MMSE receive beamforming is applied at the BS.
	\item{{\bf{Baseline with isotropic antennas}:} In this scheme, the antennas in the array are isotropic, i.e., $p = 0$ and the radiation energy is evenly distributed in the front half-space of the antennas, and the MMSE receive beamforming is applied at the BS.}
	
\end{itemize}

%\begin{figure}[!t] \centering
%	\includegraphics[width=2.5in]{fig_usersetup}
%	\caption{Simulation setup of the multi-user and multipath channel case (top view).}
%	\label{fig_usersetup}
%\end{figure}

\begin{figure}[!t]  \centering
	\includegraphics[width=2.67in]{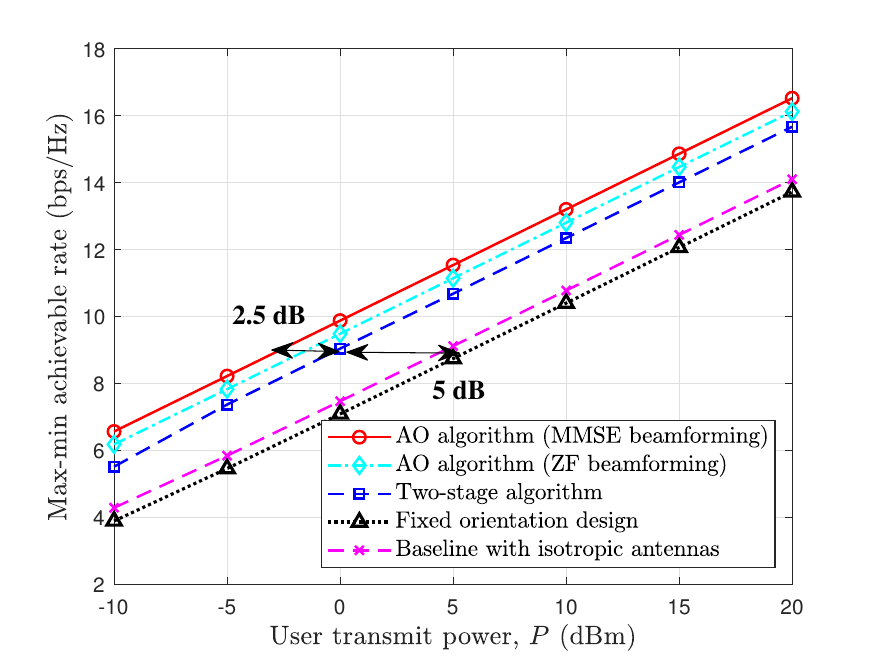}\vspace{-0.2cm}
	\caption{Max-min achievable rates obtained by different optimization algorithms versus the user transmit power $P$.}
	\label{fig_multi_alg_revise}\vspace{-0.2cm}
\end{figure}

In Fig.~\ref{fig_multi_alg_revise}, we compare the max-min achievable rates versus the user transmit power $P$ for different optimization algorithms. First, it is observed that the AO algorithm with MMSE receive beamforming achieves the highest max-min achievable rate and achieves up to 2.5 dB gain over the two-stage algorithm. As discussed in Remark~2, this gain gap is expected since the lack of iteration and the assumption of $p = 1$ in the two-stage algorithm inevitably cause performance loss. Second, by optimizing the RA pointing vectors to maximize the weighted channel power gain, the two-stage algorithm still achieves up to 5 dB gain over the fixed-antenna system. The above results indicate the different trade-offs between performance and complexity offered by the two proposed algorithms. Specifically, the AO algorithm can achieve better performance due to its iterative optimization process, while the two-stage algorithm offers competitive performance at significantly lower computational overhead.
Third, to achieve the same max-min achievable rate, the RA system requires significantly lower user transmit power than both the fixed-antenna and isotropic-antenna systems. This is because antenna rotation enables the BS to dynamically adjust its directional gain based on user distribution, thereby capturing more signal power from the users.
%These results indicate that the proposed RA architecture has strong potential to become an energy-efficient transmission technology for future green networks.
%Furthermore, since the ZF receiver enhances the noise power, the AO algorithm with MMSE beamforming always outperforms that with ZF beamforming.

\begin{figure}[!t]  \centering
	\includegraphics[width=2.67in]{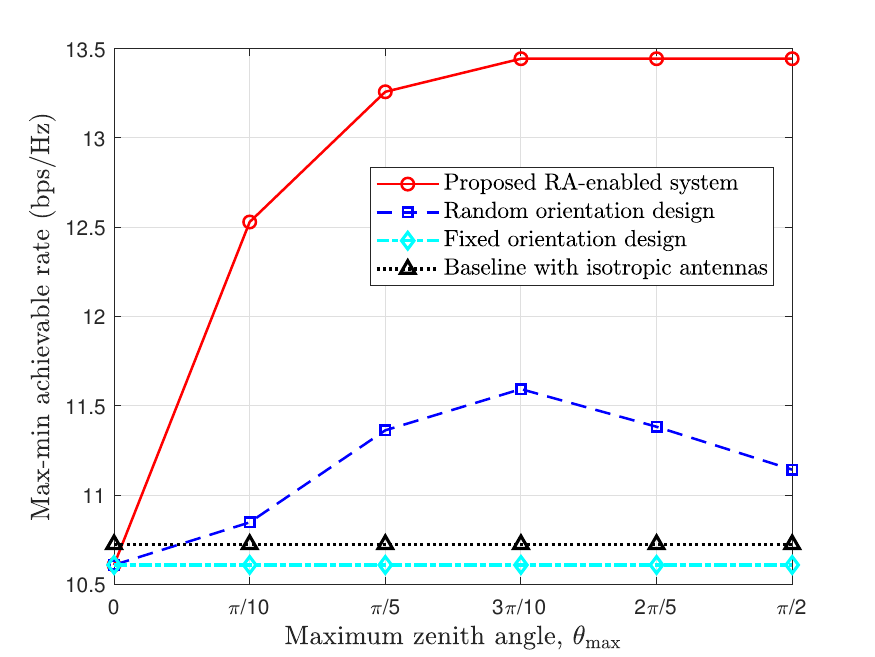}\vspace{-0.2cm}
	\caption{Max-min achievable rates of different schemes versus the maximum zenith angle $\theta_{\mathrm{max}}$.}
	\label{fig_multi_maxangle_revise}\vspace{-0.3cm}
\end{figure}

Fig.~\ref{fig_multi_maxangle_revise} shows the max-min achievable rates of different schemes versus the maximum zenith angle $\theta_{\mathrm{max}}$. Several interesting observations are made as follows. First, as $\theta_{\mathrm{max}}$ increases, the proposed RA system gains more DoFs and flexibilities to balance the array directional gain over the multipath channels, thus leading to a further increase in its max-min achievable rate. 
%Second, the proposed RA system always outperforms both the array-wise orientation adjustment counterpart and the fixed-antenna system. This is because neither of the latter two systems can independently adjust the orientation/boresight of each individual antenna to reconfigure the directional gain pattern of the entire array. 
Second, by adjusting antenna orientations/boresights to reconfigure the overall directional gain pattern, the proposed RA system consistently outperforms the fixed-antenna system.
Third, since the RAs with random orientations can statistically radiate power in any direction of the BS to serve spatially-distributed users, it can achieve a higher max-min achievable rate than the fixed-antenna system. However, when $\theta_{\mathrm{max}} \geq \frac{3\pi}{10}$, the max-min achievable rate of the random orientation design declines with $\theta_{\mathrm{max}}$. This result highlights the importance of antenna orientation/boresight optimization in an RA system, since the random orientations will lead to an unordered array directional gain pattern and inevitable performance loss when $\theta_{\mathrm{max}}$ becomes large. Last but not least, the growth rate of the max–min communication rate in our proposed RA system rises sharply when $\theta_{\mathrm{max}} \leq \frac{\pi}{10}$, showing that even a small rotational range for RA orientation/boresight adjustment can yield substantial performance gains.

\begin{figure}[!t]  \centering
	\includegraphics[width=2.67in]{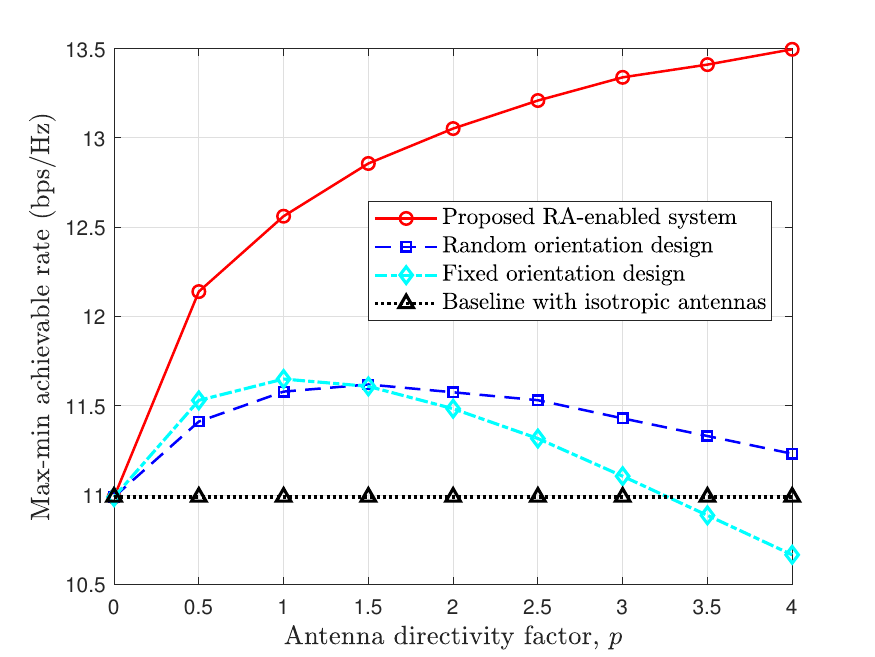}\vspace{-0.2cm}
	\caption{Max-min achievable rates of different systems versus the antenna directivity factor $p$.}
	\label{fig_multi_pattern_revise}\vspace{-0.3cm}
\end{figure}

Fig. \ref{fig_multi_pattern_revise} shows the max-min achievable rates of different schemes versus the antenna directivity factor $p$. It is observed that the max-min achievable rate of the proposed RA system increases with the directivity factor $p$. 
%This is attributed to the fact that with a larger $p$, the directional gain in the boresight direction of the antenna becomes larger and the main lobe becomes narrower, which is more advantageous for our RA system to enhance directional gains in multiple user directions by adjusting the orientations/boresights of RAs, thus resulting in a larger max-min effective communication rate.
This is attributed to the fact that a larger $p$ leads to a higher directional gain in the antenna's boresight direction and a narrower main lobe, which is more advantageous for the proposed RA system.
Consequently, the RA system with larger $p$ can more effectively enhance directional gains across multiple user directions, thereby achieving a larger max-min effective communication rate.
%{\color{blue}In contrast, the max-min achievable rate of the fixed-antenna system decreases with $p$ when $p \geq 1$. The reason is that, with a larger directivity factor $p$, the radiation power of fixed-antenna system will be more concentrated in the region directly in front of the array.}
In contrast, the max-min achievable rate of the fixed-antenna system decreases with $p$ when $p \geq 1$, as a larger directivity factor concentrates the radiated power more narrowly in front of the array.
As a result, the directional gains for users deviating from the main direction of the array will become weaker, thus resulting in a lower max-min achievable rate. Additionally, although the random orientation design can disperse the radiation power of the array in multiple directions, it is significantly inferior to the proposed RA system since it fails to strategically allocate the antenna resources to fairly improve the communication performance of all users. 
%The above results highlight the necessity of our proposed RA system for increasing channel capacity, especially in situations where the antennas have strong directivity, i.e., their main lobes are narrow.
\vspace{-0.3cm}

\subsection{Polarization-Aware Wideband Channel}
To validate the effectiveness and performance advantages of the RA architecture in practical scenarios, we extend the multi-user narrowband setting to the polarization-aware wideband channel setup, as described in Appendix~E. 
Specifically, we consider $K = 4$ users and $Q = 8$ scatterer clusters, with their geometric distributions similarly given in Section~\ref{Multi-User Narrow-Band}.
We adopt orthogonal frequency division multiplexing (OFDM) with the carrier frequency $f_c = 2.4$~GHz and bandwidth $B = 40$~MHz. The number of subcarriers and the cyclic prefix (CP) length are configured as $L = 64$ and $L_{\mathrm{CP}} = 6$, respectively. 
Each user has a transmit power budget of $P_k = 10$~dBm, and the maximum zenith angle of each RA is $\theta_{\mathrm{max}} = \frac{\pi}{6}$.
Vertically polarized antennas are employed at both the BS and users. 
%i.e., the polarized field components are given by $\tilde{F}_{\tilde{\theta}_{n,k}} = 1$ and $\tilde{F}_{\tilde{\phi}_{n,k}} = 0$. 
The achievable sum-rate defined in \eqref{deqn_ex11z} is adopted as the performance metric.
For comparison, we consider the benchmark schemes described in Section~\ref{Multi-User Narrow-Band}, applied with the same MRC beamforming and subcarrier allocation as described in Appendix~E.
%by applying the polarization-aware wideband channel model and the subcarrier allocation provided in Appendix~E, the benchmark schemes based on random orientation, fixed orientation, and isotropic antenna are still considered in this subsection.
In the polarization-unaware RA system, both the RA pointing vectors and subcarrier allocation are optimized based on the polarization-unaware channel models defined in \eqref{deqn_ex8a} and \eqref{deqn_ex9a}.

\begin{figure}[!t]  \centering
	\includegraphics[width=2.67in]{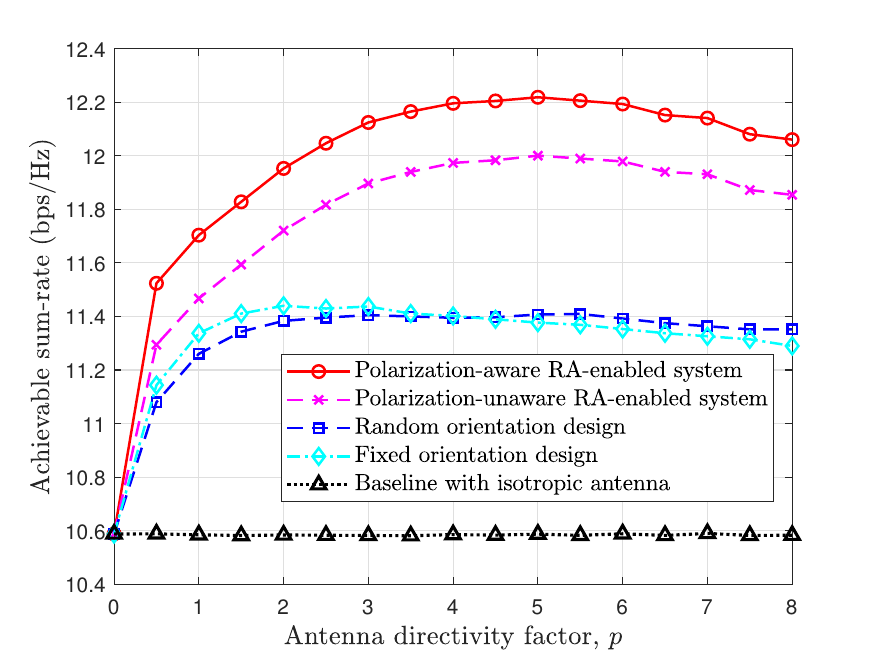}\vspace{-0.2cm}
	\caption{Achievable sum-rates of different schemes versus the antenna directivity factor $p$.}
	\label{fig_polar_direct}\vspace{-0.3cm}
\end{figure}

Fig. \ref{fig_polar_direct} shows the achievable sum-rate versus the antenna directivity factor, demonstrating the performance of RA systems under polarization-aware modeling. 
It can be observed that the achievable sum-rates of RA systems exhibit a unimodal dependence on the directivity factor $p$. Specifically, the performance improves as $p$ increases up to an intermediate value (around $p=5$), but then degrades. This behavior reflects the fundamental trade-off between radiation gain and spatial coverage: insufficient directivity fails to provide adequate gain toward the users, while excessive directivity limits the ability to capture multiple propagation paths in wideband communication.
Notably, the polarization-aware RA system consistently outperforms all other schemes across the entire range of $p$. This advantage stems from the flexible adjustment of antenna orientation/boresight direction, which effectively balances polarization alignment and directional gain to improve channel quality. These results confirm the effectiveness of the proposed RA architecture in practical wideband systems with polarization effects.
\begin{figure}[!t]  \centering
	\includegraphics[width=2.67in]{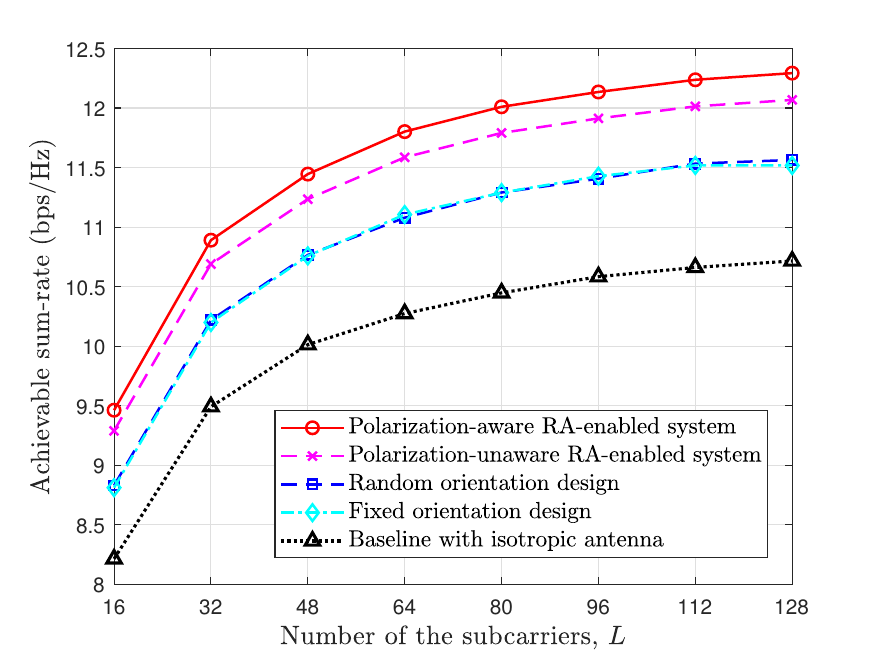}\vspace{-0.2cm}
	\caption{Achievable sum-rates of different schemes versus the number of the subcarriers $L$.}
	\label{fig_polar_carrier}\vspace{-0.3cm}
\end{figure}

Fig.~\ref{fig_polar_carrier} illustrates the achievable sum-rate versus the number of subcarriers, validating the robustness of the proposed RA design under wideband OFDM settings.
 It is observed that the achievable sum-rates of all schemes increase with the number of subcarriers. This improvement stems from higher spectral efficiency, as the relative overhead of the CP decreases with more subcarriers.
Additionally, for RA systems, increasing $L$ improves subcarrier allocation flexibility, which can be jointly optimized with antenna orientations. The polarization-aware RA system consistently achieves the best performance by jointly optimizing pointing vectors and subcarrier allocation while accounting for polarization variation due to antenna rotation. Although the polarization-unaware RA system also outperforms fixed and random orientation designs as well as the isotropic baseline, its sum-rate remains slightly lower than that of the polarization-aware counterpart. This highlights the importance of accurate polarization modeling when optimizing RA configurations. Nevertheless, the polarization-unaware design still provides a computationally efficient alternative with acceptable performance.

\vspace{-0.4cm}
\section{Conclusion}
In this paper, we proposed a new RA model that provides new spatial DoFs by flexibly adjusting the 3D orientation/boresight of each antenna. Specifically, the receive beamforming and the pointing vectors of all RAs were jointly optimized to maximize the minimum SINR among all users. The optimal closed-form pointing vectors for all RAs were first derived in the single-user and free-space propagation setup.
%Meanwhile, the asymptotic performance analysis for the case with an infinite number of antennas demonstrated that the RA system consistently achieves a higher array gain compared to the conventional fixed-antenna system.
For the general multi-user and multipath channel setup, an AO algorithm and a two-stage algorithm were proposed to obtain high-quality suboptimal solutions to balance the array directional gain over the multipath channels. Simulation results validated our analytical results and demonstrated that our proposed RA system significantly outperforms various benchmark schemes. It was shown that even with a small rotational range for RA orientation/boresight adjustment, the RA system could still reap considerable performance gains.

Building upon the general channel model and theoretical performance of the proposed RA system, future research may delve into developing new practical techniques to enhance system performance. Promising directions include the design of more efficient channel estimation algorithms and the low-overhead boresight rotation schemes, as well as the extension of the current framework to emerging 6G applications.
%Besides, the performance analysis and practical design for polarization-aware wideband RA systems are important problems requiring further investigation.

%\vspace{5cm}
%\newpage

{\appendices
\vspace{-0.4cm}
\section{Proof of Theorem 1}
First, for the case of $p = \frac{1}{2}$ and $N_x \leq \bar{N}_{x} \triangleq 2\left \lfloor  \frac{\operatorname{tan}\theta_{\mathrm{max}}}{\delta}\right \rfloor +1$, the integral in \eqref{deqn_ex4c} can be simplified as
\vspace{-0.2cm}\begin{align}
	\label{deqn_ex1f}
	\mathbb{D} = & \int_{-\frac{N_x\Delta}{2}}^{\frac{N_x\Delta}{2}}\frac{1}{1 + \frac{x^2}{r^2}} \mathrm{d}x = 2r \int_{0}^{\frac{N_x\delta}{2}} {\frac{1}{1 + x^2} \mathrm{d}x} \nonumber\\
	= & 2r \left.\operatorname{arctan}\left(x\right)\right|_{0}^{\frac{\bar{N}_{x}\delta}{2}} = 2r\operatorname{arctan}\left(\frac{\bar{N}_{x}\delta}{2}\right).
\end{align}
Then, for $N_x > \bar{N}_{x}$, the integral in \eqref{deqn_ex4c} can be calculated as
\begin{align}
	\label{deqn_ex2f}
	\hspace{-0.2cm}\mathbb{D} = & 2\hspace{-0.05cm} \left[\int_{0}^{\frac{\bar{N}_{x}\Delta}{2}}{\frac{1}{1 + \frac{x^2}{r^2}} \mathrm{d}x} \hspace{-0.05cm}+\hspace{-0.1cm} \int_{\frac{\bar{N}_{x}\Delta}{2}}^{\frac{N_x\Delta}{2}}{\frac{\operatorname{cos}\left(\operatorname{arctan}\left(|\frac{x}{r}|\right) \hspace{-0.05cm}-\hspace{-0.05cm} \theta_{\mathrm{max}}\right)}{1 + \frac{x^2}{r^2}} \mathrm{d}x}\right] \nonumber\\
	\overset{(a)}{=} & 2r \left[\int_{0}^{\frac{\bar{N}_{x}\delta}{2}} {\frac{1}{1 + x^2} \mathrm{d}x} + \int_{\frac{\bar{N}_{x}\delta}{2}}^{\frac{N_x\delta}{2}}{\frac{\operatorname{cos}\theta_{\mathrm{max}} + \operatorname{sin}\theta_{\mathrm{max}}x}{(1 + x^2)^{\frac{3}{2}}} \mathrm{d}x} \right] \nonumber\\
	\overset{(b)}{=} & 2r \left[\left.\operatorname{arctan}\left(x\right)\right|_{0}^{\frac{\bar{N}_{x}\delta}{2}} + \left.\frac{\operatorname{cos}\theta_{\mathrm{max}}x - \operatorname{sin}\theta_{\mathrm{max}}}{\sqrt{1 + x^2}}\right|_{\frac{\bar{N}_{x}\delta}{2}}^{\frac{N_x\delta}{2}} \right] \nonumber\\
	\overset{(c)}{=} & 2r \left[\theta_{\mathrm{max}} + \operatorname{sin}\left(\operatorname{arctan}\left(\frac{N_x\delta}{2}\right) - \theta_{\mathrm{max}}\right) \right],
\end{align}
where $(a)$ holds due to $\operatorname{cos}(\operatorname{arctan}(x)) = \frac{1}{\sqrt{1 + x^2}}$ and $\operatorname{sin}(\operatorname{arctan}(x)) = \frac{x}{\sqrt{1 + x^2}}$, $(b)$ follows the integral formulas 2.103.4, 2.264.5, and 2.264.6 in~\cite{Gradshteyn2007Table}, and $(c)$ holds due to the fact that $\operatorname{arctan}\left(\frac{\bar{N}_{x}\delta}{2}\right) = \theta_{\mathrm{max}}$.
Thus, based on \eqref{deqn_ex1f} and \eqref{deqn_ex2f}, the proof of Theorem~1 is completed.
\vspace{-0.4cm}
\section{Proof of Lemma 2}
For the case of $\sqrt{(N_x\Delta)^2 + (N_y\Delta)^2}\leq 2r\operatorname{tan}\theta_{\mathrm{max}}$, i.e., $\varpi_n = 0,\; \forall n$, the SNR expression in \eqref{deqn_ex3c} reduces to
\begin{align}
	\label{deqn_ex8f}
	\gamma = \frac{\bar{P}G_0 \xi \delta^2}{4\pi \Delta^2}\int_{-\frac{N_y\Delta}{2}}^{\frac{N_y\Delta}{2}}{\int_{-\frac{N_x\Delta}{2}}^{\frac{N_x\Delta}{2}}{\frac{1}{1 + \frac{1}{r^2}(x^2 + y^2)} \mathrm{d}x\mathrm{d}y }}.
\end{align}
By first integrating $x$ and then $y$, the double integral in \eqref{deqn_ex8f} can be calculated as
\begin{align}
	\label{deqn_ex9f}
	\mathbb{D} = & r^2 \int_{-\frac{N_y\delta}{2}}^{\frac{N_y\delta}{2}}{\int_{-\frac{N_x\delta}{2}}^{\frac{N_x\delta}{2}}{\frac{1}{1 + x^2 + y^2} \mathrm{d}x\mathrm{d}y}} \nonumber\\
	\overset{(d)}{=} & r^2 \int_{-\frac{N_y\delta}{2}}^{\frac{N_y\delta}{2}} \left. \frac{1}{\sqrt{1+y^2}}\operatorname{arctan}\left(\frac{x}{\sqrt{1+y^2}}\right) \mathrm{d}y \right|_{-\frac{N_x\delta}{2}}^{\frac{N_x\delta}{2}} \nonumber\\
	= & r^2 \int_{-\frac{N_y\delta}{2}}^{\frac{N_y\delta}{2}}\frac{2}{\sqrt{1+y^2}}\operatorname{arctan}\left(\frac{N_x\delta}{2\sqrt{1+y^2}}\right) \mathrm{d}y \nonumber\\
	\overset{(e)}{\approx} & \frac{1}{4}N_x\pi \delta r^2 \int_{-\frac{N_y\delta}{2}}^{\frac{N_y\delta}{2}}\frac{1}{1+y^2} \mathrm{d}y = \left.\frac{1}{4}N_x\pi \delta r^2 \operatorname{arctan}\left(y\right) \right|_{-\frac{N_y\delta}{2}}^{\frac{N_y\delta}{2}} \nonumber\\
	= & \frac{1}{2}N_x\pi \delta r^2 \operatorname{arctan}\left(\frac{N_y\delta}{2}\right) \overset{(f)}{\approx} \frac{N_x N_y \pi^2 \delta^2 r^2}{16},
\end{align}
where $(d)$ follows the integral formula 2.172 in~\cite{Gradshteyn2007Table}, $(e)$ and $(f)$ hold by exploiting the linear approximation of $\operatorname{arctan}(x) \approx \frac{\pi}{4}x$, $-1\leq x\leq 1$ \cite{Rajan2006Efficient}.
Thus, by substituting \eqref{deqn_ex9f} into \eqref{deqn_ex8f} and considering the conditions that $-1\leq \frac{N_x\delta}{2}\leq 1$ and $-1\leq \frac{N_y\delta}{2}\leq 1$, i.e., $N_x,N_y\leq \frac{2}{\delta}$ or $\theta_{\mathrm{max}} \leq \frac{\pi}{4}$, Lemma~2 can be obtained.
\vspace{-0.3cm}
\section{Proof of Theorem 2}
Based on Theorem~1 in~\cite{Feng2024Near} and the approximated SNR expression in \eqref{deqn_ex3c}, we have
\begin{align}
	\label{deqn_ex3f}
	\mathcal{F}(R_{\mathrm{lb}},p,\theta_{\mathrm{max}}) \leq \gamma \leq \mathcal{F}(R_{\mathrm{ub}},p,\theta_{\mathrm{max}}),
\end{align}
where the function $\mathcal{F}(R,p,\theta_{\mathrm{max}})$ is defined as
\begin{align}
	\label{deqn_ex4f}
	\hspace{-0.25cm}& \mathcal{F}(R,p,\theta_{\mathrm{max}}) = \frac{\bar{P} G_0 \xi\delta^2}{4\pi \Delta^2} \left(\int_{0}^{2\pi}\mathrm{d}\zeta \int_{0}^{D} \frac{1}{1 + \left(\frac{l}{r}\right)^2} l\mathrm{d}l + \right. \nonumber\\
	\hspace{-0.25cm}& \quad \quad \quad \left. \int_{0}^{2\pi}\mathrm{d}\zeta \int_{D}^{R} \frac{\operatorname{cos}^{2p}\left(\operatorname{arctan}\left(\frac{l}{r}\right) - \theta_{\mathrm{max}}\right)}{1 + \left(\frac{l}{r}\right)^2} l\mathrm{d}l \right),\hspace{-0.2cm}
\end{align}
where $D \triangleq \mathrm{min}\{R, r\operatorname{tan}\theta_{\mathrm{max}}\}$. The first double integral in \eqref{deqn_ex4f} can be calculated as
\begin{align}
	\label{deqn_ex5f}
	\mathbb{D}_1 = & 2\pi r^2 \int_{0}^{\frac{D}{r}}\frac{l}{1 + l^2}\mathrm{d}l = \pi r^2 \left. \operatorname{ln}(1+l^2)\right|_{0}^{\frac{D}{r}} \nonumber\\
	= & \pi r^2 \operatorname{ln}\left(1 + \left(\frac{D}{r}\right)^2 \right).
\end{align}
The second double integral in \eqref{deqn_ex4f} can be calculated as
\vspace{-0.2cm}\begin{align}
	\label{deqn_ex6f}
	\hspace{-0.25cm}\mathbb{D}_2 = & 2\pi r^2\int_{\frac{D}{r}}^{\frac{R}{r}}\frac{\operatorname{cos}^{2p}\left(\operatorname{arctan}\left(l\right) - \theta_{\mathrm{max}}\right)}{1 + l^2} l\mathrm{d}l \nonumber\\
	\hspace{-0.25cm}= & 2\pi r^2\int_{\operatorname{arctan}\left(\frac{D}{r}\right) - \theta_{\mathrm{max}}}^{\operatorname{arctan}\left(\frac{R}{r}\right) - \theta_{\mathrm{max}}} \operatorname{cos}^{2p}\varpi \operatorname{tan}(\varpi + \theta_{\mathrm{max}}) \mathrm{d}\varpi.\hspace{-0.2cm}
\end{align}
Thus, by substituting \eqref{deqn_ex5f} and \eqref{deqn_ex6f} into \eqref{deqn_ex3f} and \eqref{deqn_ex4f}, Theorem~2 can be obtained.
\vspace{-0.3cm}
\section{Proof of Lemma 3}
For $p = \frac{1}{2}$, the integral in \eqref{deqn_ex9c} can be calculated as
\vspace{-0.2cm}\begin{align}
	\label{deqn_ex7f}
	\mathbb{D} = & \int_{\operatorname{arctan}\left(\frac{D}{r}\right)}^{\operatorname{arctan}\left(\frac{R}{r}\right)} \operatorname{cos}(\varpi - \theta_{\mathrm{max}})\operatorname{tan}\varpi \mathrm{d}\varpi \nonumber\\
	= & \int_{\operatorname{arctan}\left(\frac{D}{r}\right)}^{\operatorname{arctan}\left(\frac{R}{r}\right)} \left(\operatorname{cos}\theta_{\mathrm{max}}\operatorname{sin}\varpi + \operatorname{sin}\theta_{\mathrm{max}} \frac{\operatorname{sin}^2\varpi}{\operatorname{cos}\varpi }\right) \mathrm{d}\varpi \nonumber\\
	= & \operatorname{cos}\theta_{\mathrm{max}}\hspace{-0.05cm} \left(\operatorname{cos}\left(\operatorname{arctan}\left(\hspace{-0.05cm}\frac{D}{r}\right)\hspace{-0.05cm}\right) \hspace{-0.05cm}-\hspace{-0.05cm} \operatorname{cos}\left(\operatorname{arctan}\left(\hspace{-0.05cm}\frac{R}{r}\hspace{-0.02cm}\right)\right)\right)\hspace{-0.05cm}+ \nonumber\\
	& \quad \operatorname{sin}\theta_{\mathrm{max}}\int_{\operatorname{sin}\left(\operatorname{arctan}\left(\frac{D}{r}\right)\right)}^{\operatorname{sin}\left(\operatorname{arctan}\left(\frac{R}{r}\right)\right)} \frac{\varpi^2}{1-\varpi^2} \mathrm{d}\varpi \nonumber\\
	\overset{(g)}{=} & \operatorname{cos}\theta_{\mathrm{max}} \left(\operatorname{cos}\theta_{\mathrm{max}} - \operatorname{cos}\theta_R\right)+ \nonumber\\
	& \quad \operatorname{sin}\theta_{\mathrm{max}}\left.\left(\frac{1}{2}\operatorname{ln}\left(\frac{1 + \varpi}{1 - \varpi}\right) - \varpi \right)\right|_{\operatorname{sin}\theta_{\mathrm{max}}}^{\operatorname{sin}\theta_R} \nonumber\\
	= & 1 - \operatorname{cos}\left(\theta_R - \theta_{\mathrm{max}} \right) + \nonumber\\
	& \quad \frac{\operatorname{sin}\theta_{\mathrm{max}}}{2} \operatorname{ln}\left(\frac{(1 + \operatorname{sin}\theta_R)(1 - \operatorname{sin}\theta_{\mathrm{max}})}{(1 - \operatorname{sin}\theta_R)(1 + \operatorname{sin}\theta_{\mathrm{max}})}\right),
\end{align}
where $\theta_R \triangleq \operatorname{arctan}\left(\frac{R}{r}\right)$, $(g)$ follows the integral formulas 2.147.1 and 2.111.6 in~\cite{Gradshteyn2007Table}, and it has $\theta_{\mathrm{max}} = \operatorname{arctan}\left(\frac{D}{r}\right)$ when $R \geq D$.
Thus, by substituting \eqref{deqn_ex7f} into \eqref{deqn_ex9c}, Lemma~3 can be obtained.
\vspace{-0.4cm}

\section{Polarization-Aware Wideband Channel Model}
Polarization is a fundamental characteristic of electromagnetic (EM) waves, and the polarization direction denotes the oscillation orientation of the electric field~\cite{Balanis1996Antenna}.
In RA systems, antenna rotation will change the polarization axis, which may lead to misalignment between transmit and receive polarizations and thereby induce polarization mismatch loss. To incorporate this effect into the RA channel model, we assume that both the BS and users employ linearly polarized antennas. As illustrated in Fig.~\ref{fig_polarization}, the electric field vector of an EM wave is always perpendicular to its propagation direction and lies in the plane formed by the antenna polarization axis (i.e., main polarization direction) and the propagation direction.
\begin{figure}[!t] 
    \centering
	\includegraphics[width=3.5in]{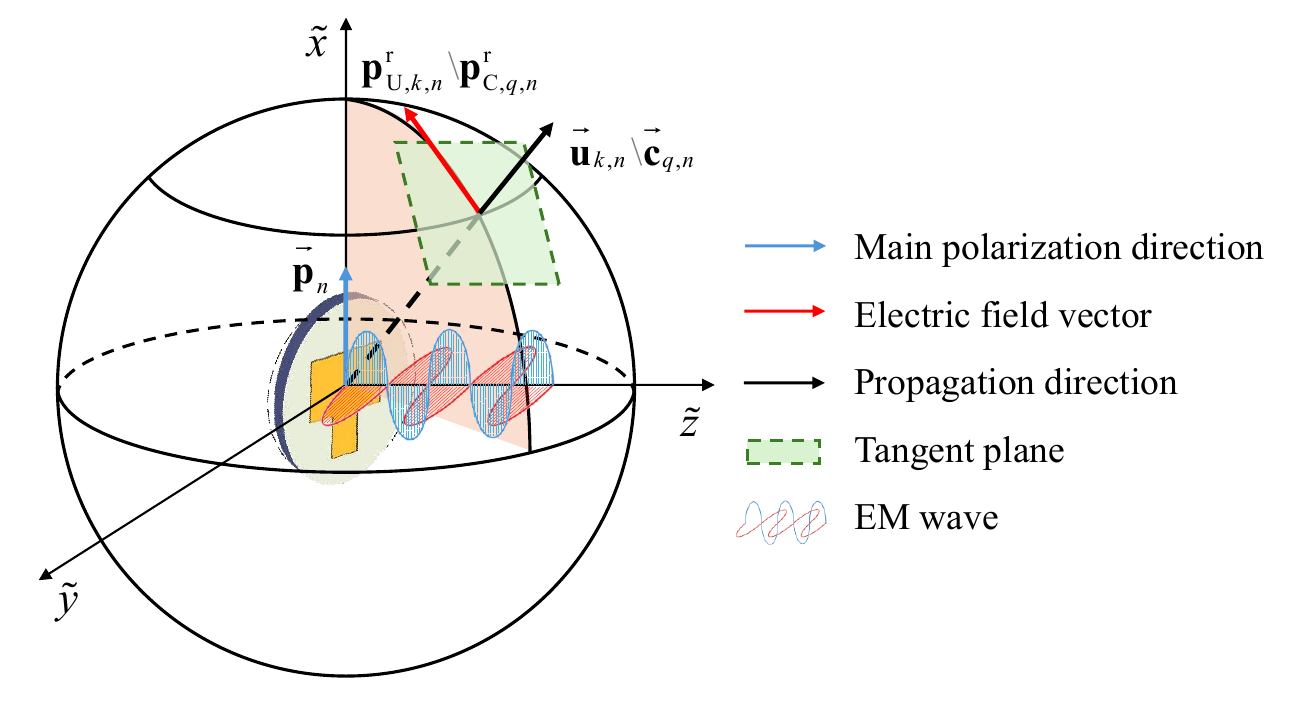} \vspace{-0.7cm}
	\caption{Illustration of polarization direction and electric field vector.}
	\label{fig_polarization}\vspace{-0.3cm}
\end{figure}

Assume that the main polarization direction of each RA is initially aligned with the $\tilde{x}$-axis in its local coordinate system $\tilde{o}$-$\tilde{x}\tilde{y}\tilde{z}$ as shown in Fig.~\ref{fig_polarization}, consistent with the pointing vector defined in \eqref{deqn_ex1a}. After antenna rotation, the polarization direction of RA~$n$ can be expressed as
\vspace{-0.2cm}
\begin{align}
	\label{deqn_ex13z}
	\vec{\mathbf{p}}_n = [\operatorname{cos}\theta_{\mathrm{z},n}\operatorname{cos}\theta_{\mathrm{a},n},\operatorname{cos}\theta_{\mathrm{z},n}\operatorname{sin}\theta_{\mathrm{a},n},-\operatorname{sin}\theta_{\mathrm{z},n}]^T.
\end{align}
The projection of $\vec{\mathbf{p}}_n$ onto the tangent plane orthogonal to the EM wave propagation from user~$k$ to RA~$n$ is given by
\vspace{-0.2cm}
\begin{align}
	\label{deqn_ex1z}
	\mathbf{p}_{\mathrm{U},k,n}^{\mathrm{r}}=\vec{\mathbf{p}}_{n} - (\vec{\mathbf{p}}_{n}^T \vec{\mathbf{u}}_{k,n})\vec{\mathbf{u}}_{k,n}.
\end{align}
Similarly, the projection of $\vec{\mathbf{p}}_n$ onto the tangent plane orthogonal to the EM wave propagation from scatterer cluster~$q$ to RA~$n$ can be obtained and is denoted by $\mathbf{p}_{\mathrm{C},q,n}^{\mathrm{r}}$.

Let $\mathbf{p}_{\mathrm{U},k,n}^{\mathrm{t}}$ and $\mathbf{p}_{\mathrm{C},q,n}^{\mathrm{t}}$ denote the electric field vectors of the EM waves transmitted from user~$k$ and reflected from scatterer cluster~$q$, respectively. The channel gains accounting for polarization effects are then given by
\vspace{-0.2cm}
\begin{align}
    \varrho_{\mathrm{U},k,n} = (\mathbf{p}_{\mathrm{U},k,n}^{\mathrm{t}})^T \mathbf{p}_{\mathrm{U},k,n}^{\mathrm{r}},\label{deqn_ex2z}\\
    \varrho_{\mathrm{C},q,n} = (\mathbf{p}_{\mathrm{C},q,n}^{\mathrm{t}})^T \mathbf{p}_{\mathrm{C},q,n}^{\mathrm{r}},\label{deqn_ex12z}
\end{align}
for the user~$k$-RA~$n$ and cluster~$q$-RA~$n$ links, respectively. 

By incorporating polarization effects, the channel gains of the LoS and NLoS links from user~$k$ to RA~$n$ are updated as
\vspace{-0.25cm}
\begin{align}
	\hspace{-0.2cm}\tilde{g}_{k, 0} (\vec{\mathbf{f}}_{n} )&=\varrho_{\mathrm{U},k,n} \sqrt{g_{\mathrm{U}, k} (\vec{\mathbf{f}}_{n} )}, \label{deqn_ex3z}\\
	\hspace{-0.2cm}{\tilde g_{k,q}}({\vec{\bf{f}} _n}) &=\frac{{\varrho_{\mathrm{C},q,n}}}{{{r_{k,q}^{\mathrm{U-C}}}}} \sqrt{{\varsigma _q}\;{g_{{\rm{C}},q}}({\vec{\bf{f}} _n})}\;e^{j\chi_q},\; q = 1,2,\ldots,Q.\hspace{-0.3cm} \label{deqn_ex4z}
\end{align}

Let ${\tau _{n,k,0}} = \frac{{{r_{k,n}^{\mathrm{B-U}}}}}{c}$ and ${\tau _{n,k,q}} = \frac{{{r_{q,n}^{\mathrm{B-C}}} + {r_{k,q}^{\mathrm{U-C}}}}}{c}$ denote the propagation delay of the ``user~$k$-RA~$n$'' and ``user~$k$-cluster~$q$-RA~$n$'' links, respectively, where $c$ is the velocity of light. Before receiver filtering and sampling, the continuous-time baseband equivalent channel impulse response between user~$k$ and RA~$n$ is given~by
\vspace{-0.25cm}
\begin{align}
	\label{deqn_ex5z}
	{\tilde h_k} ( {{{\vec{\bf{f}} }_n},t} ) = \sum\limits_{q = 0}^Q  {{{\tilde g}_{k,q}} ( {{{\vec{\bf{f}} }_n}} )}  \;{e^{ - j2\pi {f_c}{\tau _{n,k,q}}}}\delta \left( {t - {\tau _{n,k,q}}} \right),
\end{align}
where $f_c$ is the carrier frequency. The continuous-time Fourier transformation of ${\tilde h_k} ( {{{\vec{\bf{f}} }_n},t} )$ is then obtained as~\cite{Das2024Multipath}
\vspace{-0.2cm}
\begin{align}
	\label{deqn_ex6z}
	{\tilde H_k} ( {{{\vec{\bf{f}} }_n},f} ) = \sum\limits_{q = 0}^Q {{{\tilde g}_{k,q}} ( {{{\vec{\bf{f}} }_n}} )} \;{e^{ - j2\pi {f_c}\left( {1 + \frac{f}{{{f_c}}}} \right){\tau _{n,k,q}}}},
\end{align}
which denotes the spatial-frequency channel response between user~$k$ and RA~$n$. The discrete-time frequency-domain channel used for OFDM demodulation can be obtained from \eqref{deqn_ex6z} under Nyquist pulse shaping and with a sufficiently long CP.
Let $\Delta f$ denote the OFDM subcarrier spacing, and $B = L\Delta f$ be the total bandwidth with $L$ subcarriers. The frequency of the $l$-th subcarrier is ${f_l} = (l - 1)\Delta f$ with $l = 1,2,\ldots,L$~\cite{Das2024Multipath}. Therefore, spatial-frequency channel response between user~$k$ and the BS at subcarrier~$l$ is given by
\vspace{-0.2cm}
\begin{align}
	\label{deqn_ex7z}
	\hspace{-0.25cm}\tilde{\boldsymbol{h}}_{k, l}(\mathbf{F})\hspace{-0.1cm}=\hspace{-0.1cm}\left[\tilde{H}_{k} (\vec{\mathbf{f}}_{1}, \hspace{-0.05cm}f_{l} ), \tilde{H}_{k} (\vec{\mathbf{f}}_{2},\hspace{-0.05cm} f_{l} ), \ldots, \tilde{H}_{k} (\vec{\mathbf{f}}_{N},\hspace{-0.05cm} f_{l} ) \right]^{T} \hspace{-0.2cm}\in\hspace{-0.1cm} \mathbb{C}^{N \hspace{-0.05cm}\times\hspace{-0.05cm} 1}.\hspace{-0.2cm}
\end{align}

To avoid inter-user interference, we assume each subcarrier is allocated to at most one user. Define a binary variable ${a_{k,l}}$, which indicates whether subcarrier~$l$ is allocated to user~$k$, i.e., ${a_{k,l}} = 1$ if subcarrier~$l$ is assigned to user~$k$, and ${a_{k,l}} = 0$ otherwise. The subcarrier allocation constraints are given by 
\vspace{-0.2cm}
\begin{align}
	\label{deqn_ex8z}
	\sum\limits_{k = 1}^K {{a_{k,l}}}  \le 1,\;\forall l,
\end{align}
and the transmit power constraint for user~$k$ is
\vspace{-0.2cm}
\begin{align}
	\label{deqn_ex9z}
	\sum\limits_{l = 1}^L {{a_{k,l}}} {P_{k,l}} \le {P_k},\;\forall k,
\end{align}
where $P_{k,l}$ denotes the transmit power of user~$k$ at subcarrier~$l$.}

If ${a_{k,l}} = 1$ and the BS has perfect synchronization (including symbol timing, frame alignment, and carrier frequency offset (CFO) compensation),  the signal transmitted from user~$k$ and received at the BS over subcarrier $l$ can be expressed as
\begin{align}
	\label{deqn_ex10z}
	{\tilde {\bf{y}}_{k,l}} = {\tilde {\bf{h}}_{k,l}}\left( {\bf{F}} \right)\sqrt {{P_{k,l}}} {s_{k,l}} + {{\bf{n}}_l},
\end{align}
where $s_{k,l}$ denotes the information-bearing signal transmitted by user~$k$ at subcarrier~$l$, and $\mathbf{n}_l$ is the AWGN vector with variance $\tilde{\sigma}_l^2$. Since there is no inter-user interference over each subcarrier, the BS can apply the MRC beamforming, i.e., ${\tilde {\bf{v}}_{k,l}} = \frac{{{{\tilde {\bf{h}}}_{k,l}}\left( {\bf{F}} \right)}}{{ \| {{{\tilde {\bf{h}}}_{k,l}}\left( {\bf{F}} \right)} \|}}$, to maximize the receive SNR. Thus, the achievable rate in bits per second per Hertz (bps/Hz) of user~$k$ is given by
\vspace{-0.2cm}
\begin{align}
	\label{deqn_ex11z}
	{\tilde R_k} = \frac{1}{{L + {L_{{\rm{CP}}}}}}\sum\limits_{l = 1}^L {{a_{k,l}}{{\log }_2}\left( {1 + \frac{{{P_{k,l}}{{\| {{{\tilde {\bf{h}}}_{k,l}}\left( {\bf{F}} \right)} \|}^2}}}{{\tilde{\sigma}_l^2}}} \right)},
\end{align}
where $L_{{\rm{CP}}}$ is the length of the CP.

We formulate the following optimization problem to maximize the sum rate across all users and subcarriers by jointly optimizing the RA pointing matrix $\mathbf{F}$ and the subcarrier allocation $\{a_{k,l}\}$.
\vspace{-0.4cm}
\begin{subequations}\label{eq:11}
	\begin{alignat}{2}
		\text{(P11):} \quad \max_{\mathbf{F},\{a_{k,l}\}} \quad & \sum_{k=1}^{K} \tilde{R}_k & \label{eq:11A}\\
		\mbox{s.t.} \quad 
		& a_{k,l}\in\{0,1\},\; \forall k,l, \label{eq:11B}\\
		& \eqref{eq:1B}, \eqref{eq:1C}, \eqref{deqn_ex8z}, \eqref{deqn_ex9z}. \label{eq:11C}
	\end{alignat}
\end{subequations}

To solve the above problem, we can alternately optimize the pointing matrix $\mathbf{F}$ and the subcarrier allocation $\{a_{k,l}\}$ in an iterative manner. For a given $\mathbf{F}$, the subproblem of optimizing $\{a_{k,l}\}$ is a typical binary integer programming problem, which can be effectively solved using the CVX solver~\cite{Boyd2004Convex}. However, for a given $\{a_{k,l}\}$, the subproblem of optimizing $\mathbf{F}$ remains highly non-convex. In addition, the objective function \eqref{eq:11A} is non-concave, and constraint \eqref{eq:1C} is non-convex. To overcome these challenges, we first linearly approximate equality constraint \eqref{eq:1C} to approximate the feasible region as a convex set. Then, the conditional gradient method is employed to obtain the solution of $\mathbf{F}$ by using a linear programming solver, such as linprog~\cite{Rocha2019A}. Based on the principle of the BCD method, the optimized $\mathbf{F}$ and $\{a_{k,l}\}$ can be obtained by alternatively solving the above two subproblems until convergence is attained.

\vspace{-0.5cm}
\bibliographystyle{IEEEtran}
\bibliography{RA_Modeling}

\begin{IEEEbiography}[{\includegraphics[width=1in,height=1.25in,clip,keepaspectratio]{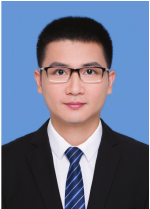}}]{Beixiong Zheng}
(Senior Member, IEEE) received his B.Eng. and Ph.D. degrees from the South China University of Technology, Guangzhou, China, in 2013 and 2018, respectively. 
He is currently a Professor with the School of Microelectronics, South China University of Technology. He was a Research Fellow with the Department of Electrical and Computer Engineering, National University of Singapore from 2019 to 2022. From 2015 to 2016, he was a Visiting Student Research Collaborator with Columbia University, New York, NY, USA. 
His recent research interests include 5G/6G wireless communications, rotatable antenna (RA), intelligent reflecting surface (IRS), and signal processing.

Dr. Zheng is now serving as the Lead Guest Editor for the \textsc{IEEE Journal on Selected Areas in Communications (JSAC)}, 
and an Editor for the \textsc{IEEE Transactions on Wireless Communications (TWC)} and
the \textsc{IEEE Communications Letters (CL)}.
He was the recipient of
the \textsc{IEEE Communications Society} Asia-Pacific Best Young Researcher Award in 2025,
the \textsc{IEEE Communications Society} Heinrich Hertz Award for Best Communications Letter in 2022,
the \textsc{IEEE Communications Society} Best Tutorial Paper Award in 2023,
the Electronic Information Science and Technology Award (First Prize of Natural Science Award) of Guangdong Province in 2022,
the Science and Technology Award (First Prize of Natural Science Award) of Guangdong Communications Society in 2024,
the Best Ph.D. Thesis Award from the China Education Society of Electronics in 2018,
the Best Demo Award from the IEEE/CIC International Conference on Communications in China (ICCC) in 2025,
the Best Paper Award from the IEEE International Conference on Computing, Networking and Communications (ICNC) in 2016,
the Best Paper Award from the IEEE International Conference on Wireless Communications and Signal Processing (WCSP) in 2024,
and the Best Paper Award from the International Conference on Ubiquitous Communication (Ucom)  in 2023.
He was also awarded the Humboldt Research Fellowship for experienced researchers in 2024.
In addition, he was listed as the Highly Cited Researchers by Clarivate,
the Highly Cited Chinese Researchers by Elsevier, and 
the World’s Top 2\% Scientist by Stanford University and Mendeley Data. 
%He was also recognized as an Exemplary Reviewer of the \textsc{IEEE Transactions on Communications} and \textsc{IEEE Communications Letters}, and an Outstanding Reviewer of the \textsc{Physical Communication}.
%He was also the recipient of the Exemplary Reviewer of the \textsc{IEEE Transactions on Communications} and \textsc{IEEE Communications Letters}, and the Outstanding Reviewer of the \textsc{Physical Communication}.
\end{IEEEbiography}

\begin{IEEEbiography}[{\includegraphics[width=1in,height=1.25in,clip,keepaspectratio]{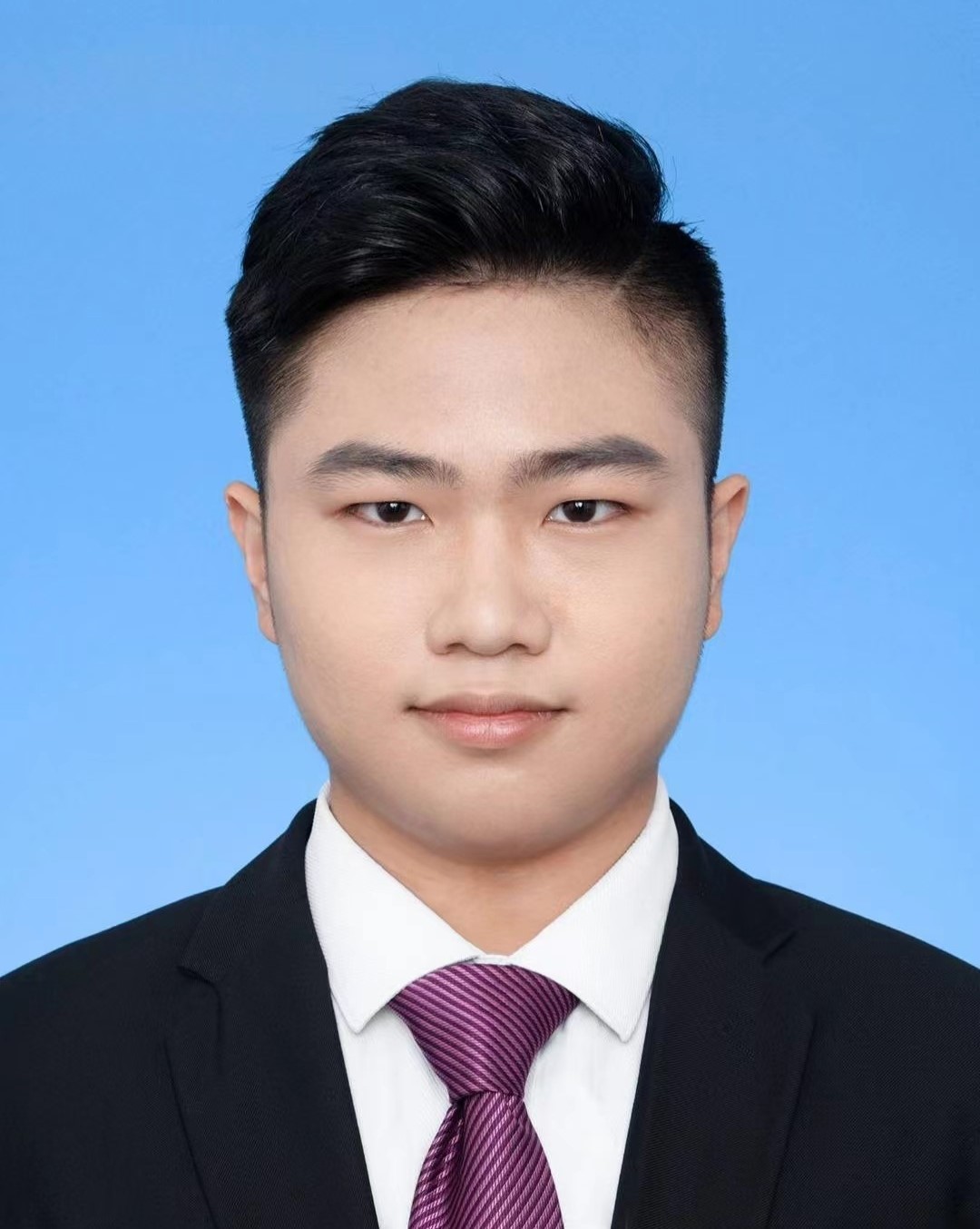}}]{Qingjie Wu} received the B.Eng. degree in communications engineering and the M.Eng. degree in information and communication engineering from Guangdong University of Technology, Guangzhou, China, in 2021 and 2024, respectively. He is currently pursuing the Ph.D. degree in electronic science and technology with the School of Microelectronics, South China University of Technology. His research interests include rotatable antenna (RA)-enabled wireless communications, unmanned aerial vehicle (UAV) communications, intelligent reflecting surface (IRS), ultra-reliable and low-latency communications (URLLC), and convex optimization.
\end{IEEEbiography}

\begin{IEEEbiography}[{\includegraphics[width=1in, height=1.25in,clip, keepaspectratio]{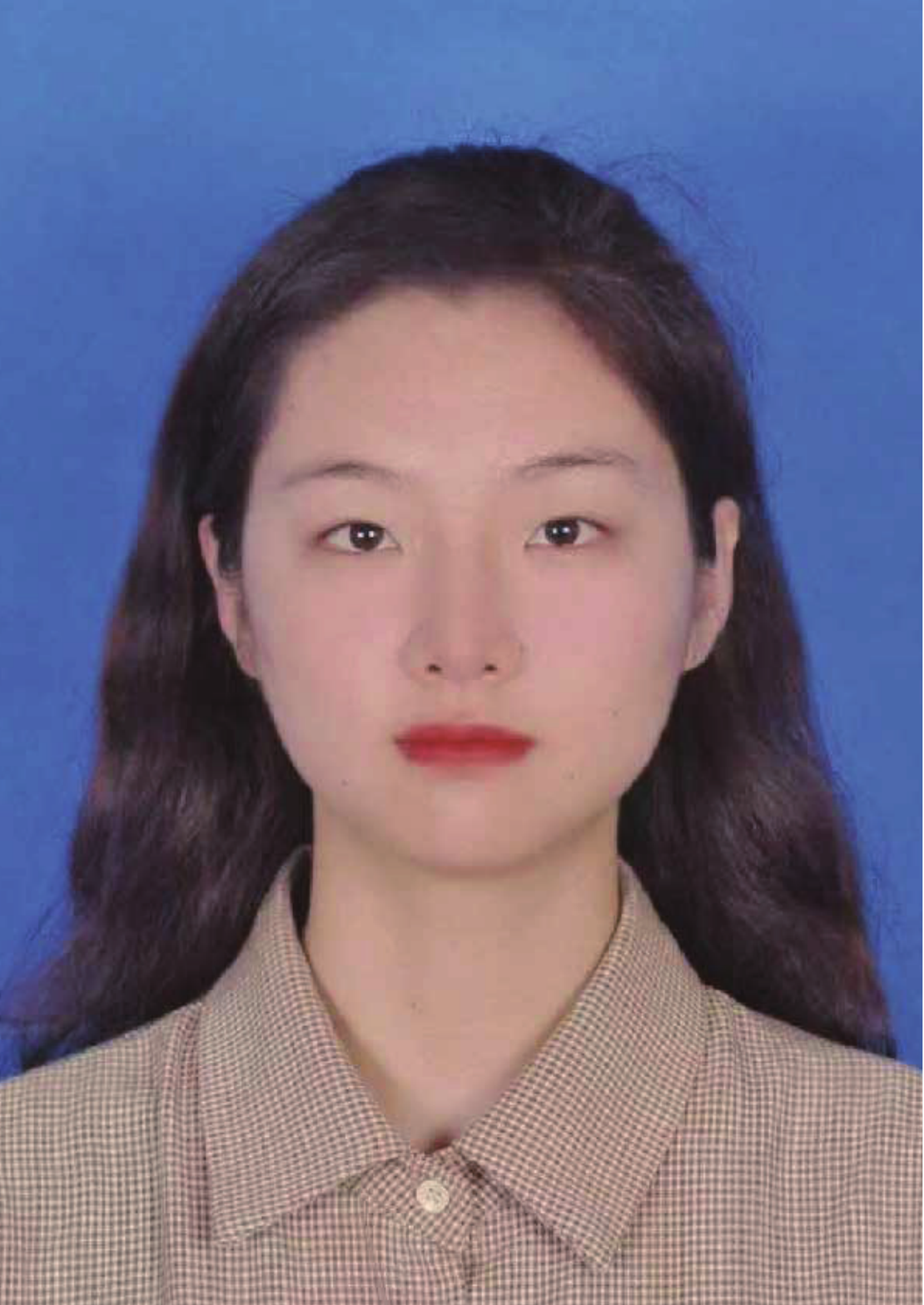}}]{Tiantian Ma} received the B.E. degree from the South China University of Technology, Guangzhou, China, in 2023, where she is currently pursuing the Ph.D. degree in electronic science and technology. Her research interests include 5G/6G wireless communications, rotatable antenna (RA), and signal processing. 
\end{IEEEbiography}
\begin{IEEEbiography}[{\includegraphics[width=1.4in,height=1.3in,keepaspectratio]{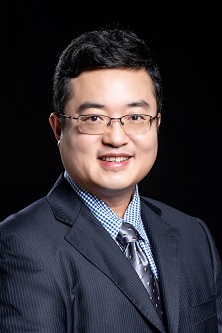}}] {Rui Zhang} (S'00-M'07-SM'15-F'17) received the B.Eng. (first-class Hons.) and M.Eng. degrees from the National University of Singapore, Singapore, and the Ph.D. degree from the Stanford University, Stanford, CA, USA, all in electrical engineering.
 
From 2007 to 2009, he worked as a research scientist at the Institute for Infocomm Research, ASTAR, Singapore. In 2010, he joined the Department of Electrical and Computer Engineering of National University of Singapore, where he is now a Provost’s Chair Professor. He is also an Adjunct Professor with the School of Science and Engineering, The Chinese University of Hong Kong, Shenzhen, China. He has published over 600 papers, all in the field of wireless communications and networks. He has been listed as a Highly Cited Researcher by Thomson Reuters/Clarivate Analytics since 2015. His current research interests include intelligent surfaces, reconfigurable antennas, radio mapping, non-terrestrial communications, wireless power transfer, AI and optimization methods.     
He was the recipient of the 6th IEEE Communications Society Asia-Pacific Region Best Young Researcher Award in 2011, the Young Researcher Award of National University of Singapore in 2015, the Wireless Communications Technical Committee Recognition Award in 2020, the IEEE Signal Processing and Computing for Communications (SPCC) Technical Recognition Award in 2021, the IEEE Communications Society Technical Committee on Cognitive Networks (TCCN) Recognition Award in 2023, and the IEEE James Evans Avant Garde Award in 2025. His works received 18 IEEE Best Journal Paper Awards, including the IEEE Marconi Prize Paper Award in Wireless Communications in 2015 and 2020, the IEEE Signal Processing Society Best Paper Award in 2016, the IEEE Communications Society Heinrich Hertz Prize Paper Award in 2017, 2020 and 2022, the IEEE Communications Society Stephen O. Rice Prize in 2021, etc. He served for over 30 international conferences as the TPC co-chair or an organizing committee member. He was an elected member of the IEEE Signal Processing Society SPCOM Technical Committee from 2012 to 2017 and SAM Technical Committee from 2013 to 2015. He served as the Vice Chair of the IEEE Communications Society Asia-Pacific Board Technical Affairs Committee from 2014 to 2015, a member of the Steering Committee of the IEEE Wireless Communications Letters from 2018 to 2021, a member of the IEEE Communications Society Wireless Communications Technical Committee (WTC) Award Committee from 2023 to 2025. He was a Distinguished Lecturer of IEEE Signal Processing Society and IEEE Communications Society from 2019 to 2020. He served as an Editor for several IEEE journals, including the IEEE TRANSACTIONS ON WIRELESS COMMUNICATIONS from 2012 to 2016, the IEEE JOURNAL ON SELECTED AREAS IN COMMUNICATIONS: Green Communications and Networking Series from 2015 to 2016, the IEEE TRANSACTIONS ON SIGNAL PROCESSING from 2013 to 2017, the IEEE TRANSACTIONS ON GREEN COMMUNICATIONS AND NETWORKING from 2016 to 2020, and the IEEE TRANSACTIONS ON COMMUNICATIONS from 2017 to 2022. He now serves as an Editorial Board Member of npj Wireless Technology, and the Chair of the IEEE Communications Society Wireless Communications Technical Committee (WTC) Award Committee. He is a Fellow of the Academy of Engineering Singapore. \end{IEEEbiography}

\end{document}